\documentclass[aps,pre,superscriptaddress,twocolumn,longbibliography,groupedaddress]{revtex4-2}

\usepackage[colorlinks]{hyperref}
\hypersetup{
  linkcolor = {blue},
  citecolor = {blue},
  urlcolor = {cyan}
}

\usepackage{float}
\usepackage{graphicx}
\usepackage{caption}
\usepackage{subcaption}
\usepackage{amsmath}
\usepackage{amsthm,amssymb}
\usepackage{amsfonts}
\usepackage{dcolumn}
\usepackage{bm}
\usepackage{epstopdf}
\usepackage{algorithm}
\usepackage{algpseudocode}

\usepackage{epsfig}
\usepackage{mathrsfs}
\usepackage{multirow}
\usepackage{tabularx}
\usepackage{mathptmx}%
\usepackage[all]{xy}
\usepackage{pbox}
\usepackage{verbatim}
\usepackage{mathtools}
\usepackage{tikz,pgfplots}
\usepackage{xfrac}
\usepackage{cleveref}
\usepackage{booktabs}
\usepackage{physics}
\usepackage{siunitx}
\usepackage{url}
\usepackage{pict2e}
\usepackage{tikz,tikz-3dplot}
\usetikzlibrary{perspective}
\usepackage{xkeyval}
\usepackage{xcolor}
\usepackage[T1]{fontenc}
\usepackage{halloweenmath}
\usepackage{curve2e}
\usepackage{xtab,afterpage}
\usepackage{balance}
\usepackage{soul}

\newcommand{\opcatMx}[1]{{#1}_{\mathrm{x}}}
\newcommand{\opcatMz}[1]{{#1}_{\mathrm{z}}}

\newcommand{\opcatMzzzz}[1]{{#1}_{\mathrm{zzzz}}}

\newcommand{\opcatSPyM}[1]{{#1}_{\mathrm{SPyM}}}
\newcommand{\opcatBW}[1]{{#1}_{\mathrm{BW}}}

\newcommand{\downtriangle}{%
\begin{tikzpicture}[line width={0.05em},scale=0.015\baselineskip]
  \draw (0,0) -- (1,0) -- (1/2,-1) -- cycle;%
\end{tikzpicture}%
}%

\newcommand{\uppertriangle}{%
\begin{tikzpicture}[line width={0.05em},scale=0.015\baselineskip]
  \draw (0,0) -- (1,0) -- (1/2, 1) -- cycle;%
\end{tikzpicture}%
}%

\makeatletter

\@ifdefinable\SuCmathpictvertex{} 
\@ifdefinable\@SuC@reserved@dimen{\newdimen\@SuC@reserved@dimen}

\newenvironment*{@SuC@math@picture}[8]{%
  \def\SuCmathpictvertex{\circle*{#6}}%
  \setlength\unitlength{\fontdimen 22 #5\tw@}%
  \setlength\@SuC@reserved@dimen{#7\unitlength}%
  \kern\@SuC@reserved@dimen
  \@HwM@d@pict@strut{#2}%
  \picture(#3,#1)(#4,-1)%
    \roundcap
    \roundjoin
    \linethickness{#8\@HwM@thickness@units@for #5}%
}{%
  \endpicture
  \kern\@SuC@reserved@dimen
}
\newcommand*\@SuC@general@pict[9]{%
  \begin{@SuC@math@picture}%
            {#2}{#3}
            {#4}{#5}
            #6
            {#7}
            {#8}
            {#9}
    #1%
  \end{@SuC@math@picture}%
}
\newcommand*\@SuC@math@version@shunt[7]{%
  \@HwM@choose@thicknesses{\@SuC@general@pict {#1}{#2}{#3}{#4}{#5}#7}%
      {{.8}{.4}{}}
      {{1}{.5}{1.5}}
}

\newcommand*\DeclareNewSuCMathPict[6]{%
  \newcommand*{#1}{%
    \@HwM@general@ordinary@symbol
      {\@SuC@math@version@shunt {#6}{#2}{#3}{#4}{#5}}%
  }%
}

\makeatother

\DeclareNewSuCMathPict{\trigonleft}
            {3}{1}  
            {4}{-2} 
{
    \polygon(45:2)(135:2)(225:2)%
    \put (45:2){\SuCmathpictvertex}%
    \put(135:2){\SuCmathpictvertex}%
    \put(225:2){\SuCmathpictvertex}%
}

\DeclareNewSuCMathPict{\trigonright}
            {3}{1}  
            {4}{-2} 
{
    \polygon(45:2)(135:2)(315:2)%
    \put (45:2){\SuCmathpictvertex}%
    \put(135:2){\SuCmathpictvertex}%
    \put(315:2){\SuCmathpictvertex}%
}

\DeclareNewSuCMathPict{\trigonmiddle}
            {3}{1}  
            {4}{-2} 
{
    \polygon(45:2)(135:2)(270:2)%
    \put (45:2){\SuCmathpictvertex}%
    \put(135:2){\SuCmathpictvertex}%
    \put(270:2){\SuCmathpictvertex}%
}

\DeclareNewSuCMathPict{\tetrapleuro}
            {3}{1}  
            {4}{-2} 
{
    \polygon(45:2)(135:2)(270:2)%
    \put (45:2){\SuCmathpictvertex}%
    \put (90:1.4){\SuCmathpictvertex}%
    \put(135:2){\SuCmathpictvertex}%
    \put(270:2){\SuCmathpictvertex}%
}

\DeclareNewSuCMathPict{\pyramid}
            {3}{1}  
            {4}{-2} 
{
    \polygon(290:2)(40:2)(110:2)(10:2)%
}

\DeclareNewSuCMathPict{\Isingdiagonalsecond}
            {3}{1}  
            {4}{-2} 
{
    \polygon(45:2)(225:2)%
    \put (45:2){\SuCmathpictvertex}%
    \put(225:2){\SuCmathpictvertex}%
}

\DeclareNewSuCMathPict{\Isingdiagonal}
            {3}{1}  
            {4}{-2} 
{
    \polygon(135:2)(315:2)%
    \put (135:2){\SuCmathpictvertex}%
    \put(315:2){\SuCmathpictvertex}%
}

\DeclareNewSuCMathPict{\Isingvertical}
            {3}{1}  
            {4}{-2} 
{
    \polygon(135:2)(225:2)%
    \put (135:2){\SuCmathpictvertex}%
    \put(225:2){\SuCmathpictvertex}%
}

\newlength\figureheight 
\newlength\figurewidth 

\begin{document}

\title{Cellular automata in $d$ dimensions and ground states of spin models in $(d+1)$ dimensions}

\author{Konstantinos Sfairopoulos}
\email{ksfairopoulos@gmail.com}
\affiliation{School of Physics and Astronomy, University of Nottingham, Nottingham, NG7 2RD, UK}
\affiliation{Centre for the Mathematics and Theoretical Physics of Quantum Non-Equilibrium Systems,
University of Nottingham, Nottingham, NG7 2RD, UK}
\author{Luke Causer}
\affiliation{School of Physics and Astronomy, University of Nottingham, Nottingham, NG7 2RD, UK}
\affiliation{Centre for the Mathematics and Theoretical Physics of Quantum Non-Equilibrium Systems,
University of Nottingham, Nottingham, NG7 2RD, UK}
\author{Jamie F. Mair}
\affiliation{School of Physics and Astronomy, University of Nottingham, Nottingham, NG7 2RD, UK}
\affiliation{Centre for the Mathematics and Theoretical Physics of Quantum Non-Equilibrium Systems,
University of Nottingham, Nottingham, NG7 2RD, UK}
\author{Juan P. Garrahan}
\affiliation{School of Physics and Astronomy, University of Nottingham, Nottingham, NG7 2RD, UK}
\affiliation{Centre for the Mathematics and Theoretical Physics of Quantum Non-Equilibrium Systems,
University of Nottingham, Nottingham, NG7 2RD, UK}

\begin{abstract}
We show how the trajectories of $d$-dimensional cellular automata (CA) can be used to systematically construct the ground states of $(d+1)$-dimensional classical spin models, and we characterize their quantum phase transition, when in the presence of a transverse magnetic field.
For each of the 256 one-dimensional elementary CA we 
explicitly construct the simplest local  two-dimensional classical spin model associated to the given CA, and we also describe this method for $d>1$ through selected examples. 
We illustrate our general observations with 
detailed studies of: (i) the $d=1$ CA Rule 150 and its $d=2$ four-body plaquette spin model, (ii) the $d=2$ CA whose associated model is the $d=3$ square-pyramid plaquette model, and (iii) two counter-propagating $d=1$ Rule 60 CA that correspond to the two-dimensional Baxter-Wu spin model. For the quantum spin models, we show that the connection to CAs implies a sensitivity on the approach to the thermodynamic limit via finite size scaling for their quantum phase transitions. 
\end{abstract}

\maketitle

\section{Introduction}

In this paper we provide a complete classification of the ground states of a set of two-dimensional classical spin models, through the use of the trajectories of their associated one-dimensional cellular automata (CA), irrespective of their boundary conditions.
Our method for the characterization of the ground states extends to spins in arbitrary local spin dimensions and it can also be straightforwardly generalized to higher dimensions to connect $d$-dimensional CA to $(d+1)$-dimensional spin systems, and to various plaquette interaction terms.

This method is based on reverse-engineering the Hamiltonian of the spin model from the knowledge of its ground states; that is, we obtain ``parent'' Hamiltonians for each given CA rule \cite{auerbach_interacting_1994}. The relation of the Hamiltonians to the ground states is many to one, so for each class we focus on describing the simplest Hamiltonian per ground state space. 
In one dimension we restrict ourselves to elementary CA \cite{wolfram1983statistical,martin1984algebraic}, and characterize all 256 ground state spaces and the corresponding 256 spin models. This method generalizes naturally to higher dimensional CA, which we illustrate with selected 2D examples that give rise to three-dimensional spin models. In this work, we focus on periodic boundary conditions (PBC) which give rise to the most interesting interplay between the classical ground state degeneracy for given system sizes, but we also discuss other boundary conditions.

By the addition of a transverse field term, these quantum spin models would exhibit a ground state quantum phase transition controlled by the strength of the transverse field. 
We show that the knowledge of the CA that
determines the classical ground states
provides enough information together with the various numerical simulations that we use
for identifying the characteristics of these quantum phase transitions. This behaviour is similar to what occurs in Ref.~\cite{sfairopoulos2023boundary} 
for the quantum triangular plaquette  
model (TPM) \cite{vasiloiu2020trajectory,zhou2021fractal,inack2022neural,myerson-jain2022multicritical,myerson-jain2022pascals,wiedmann2023absence}, whose classical limit \cite{newman1999glassy,garrahan2001glassy,garrahan2002glassiness,yoshida2013exotic,turner2015overlap} obeys CA Rule 60.

While we study the whole class of 2D models arising from 1D elementary CA, we focus especially on Rule 150 \cite{wolfram1983statistical,martin1984algebraic}, whose associated model was studied with an emphasis on error-correction in Ref.~\cite{nixon2021correcting}, and its quantum version, the quantum Fibonacci model, was also studied in Refs.~\cite{devakul2019fractal,you2018subsystem,2020_Tantivasadakarn} (see also Ref.~\cite{stephen2022fractionalization}). 
To illustrate how the method generalizes to higher dimensions, we consider the square pyramid model (SPyM) 
\cite{turner2015overlap,jack2016phase} and we describe the properties of its quantum phase transition. Finally, we employ our methodology to the classification of ground states and the quantum phase transition for the quantum Baxter-Wu (qBW) model \cite{capponi2014baxter-wu} (which relates to two counter-propagating CA). 

As a result, we concentrate on models with linear constraints, while the study of models with nonlinear constraints is the subject of separate paper \cite{sfairopoulos2025spinmodelsnonlinearcellular}. These simple linear constraints can also be viewed as instances of the k-XORSAT constraint satisfaction problems (CSPs) \cite{mezard2009information}. Specifically, Rule 150 consists of a 4-regular 4-XORSAT, the SPyM of a 5-regular 5-XORSAT, and the BW model of a 6-regular 3-XORSAT instance. In contrast, the rules with nonlinear constraints would be defined by general SAT constraints \cite{sfairopoulos2025spinmodelsnonlinearcellular}. Since all the models studied here are XORSAT instances, Gaussian elimination can be used for the efficient solution of their ground state space in linear-time complexity \cite{mezard2009information}. However, here we will demonstrate the use of the CA for probing their ground state spaces, despite their scaling, even with the most efficient techniques \cite{cattell2000fast,sawada2017practical}, being exponential. The reason for our choice lies in the intuitiveness of the method and its flexibility in constructing local spin models with linear (this paper), but also nonlinear constraints \cite{sfairopoulos2025spinmodelsnonlinearcellular} where Gaussian elimination is inapplicable. Our approach can be alternatively thought of as an optimization of a set of cost functions based on Boolean logic, recast as a Hamiltonian problem, with potential applications on adiabatic quantum computation \cite{whitfield2012ground-state,lucas2014ising,kohler2022translationally}. For the nonlinear models \cite{sfairopoulos2025spinmodelsnonlinearcellular}, an ordering pattern (order-by-disorder, or ObD) arises for small transverse field values which can be described by low-order degenerate perturbation theory calculations. The CA construction \cite{sfairopoulos2025spinmodelsnonlinearcellular} allows the explicit determination of the classical zero-temperature ground state space of these frustrated models, on top of which the ObD occurs.

The rest of the paper is organized as follows. 
In Sec.~\ref{CA} we discuss the dynamics of 1D and 2D CA and the properties of their limit cycles. In Sec.~\ref{Section_classicalmodels} we describe the $(d+1)$-dimensional classical spin models that can be derived from the $d$-dimensional CA. In Sec.~\ref{Quantum models} we consider the corresponding quantum spin models and their ground state phase transitions. In Sec.~\ref{conclusions} we give our conclusions. Extra results are presented in the Appendices, including the list of classical spin models emerging from all the elementary CA in Appx.~\ref{appendix:all-models}, other spin models related to non-elementary CA in Appx.~\ref{appendix:other-CA}, further details of numerical simulations for system sizes with open boundary conditions (OBC) in Appx.~\ref{appendix:numerics} and a collection of the respective low energy spectra for some system sizes for Rule 150 in Appx.~\ref{appendix:low-lying}. 
\section{Cellular Automata and their Attractor Structure}{\label{CA}}

\subsection{1D Cellular Automata}{\label{1d_CA}}

Cellular automata describe the discrete-time evolution of an array of sites (or {\em cells}), belonging to a finite field or its generalizations, which are characterized by a dynamical map (or the {\em update rule}), and, in general, might not be deterministic or Markovian \cite{wolfram1983statistical,martin1984algebraic}. 
While our methods can be applied to CA with any neighbourhood and with elements in any finite field, we will focus on deterministic update rules and elementary CA. 

\begin{figure*}
    \centering
    \begin{subfigure}[b]{0.35\textwidth}
        \includegraphics[width=\textwidth, height=20mm]{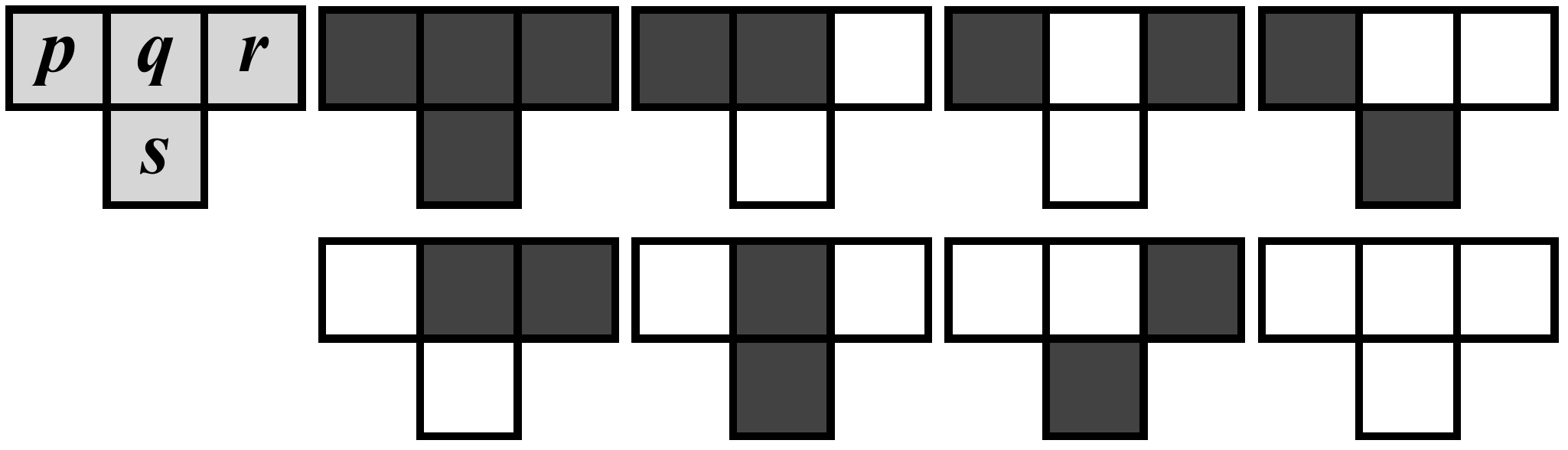}
        \caption{}
    \end{subfigure}
    \begin{subfigure}[b]{0.3\textwidth}
        \includegraphics[width=\textwidth, height=30mm]{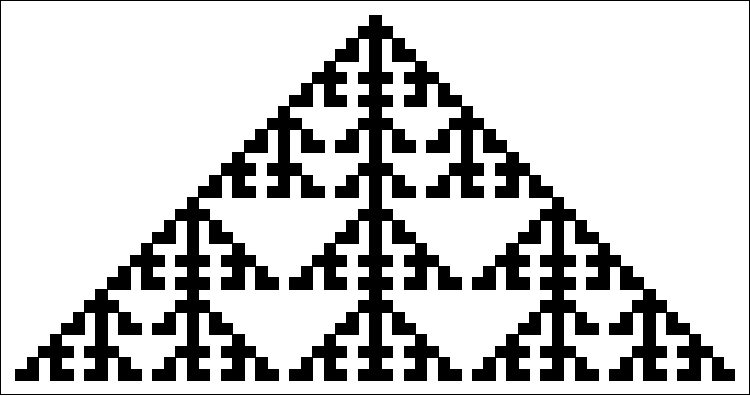}
        \caption{}
    \end{subfigure}
    \begin{subfigure}[b]{0.3\textwidth}
        \includegraphics[width=\textwidth, height=30mm]{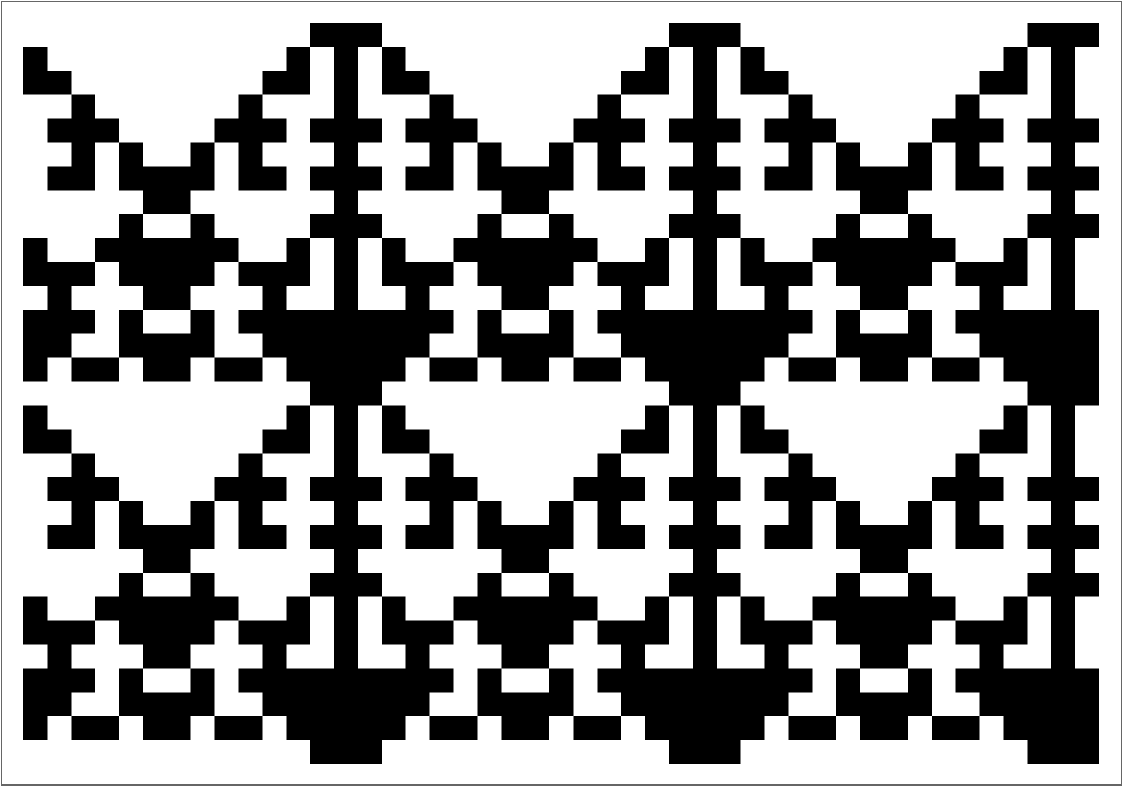}
        \caption{}
    \end{subfigure}
    \caption{
        {\bf CA Rule 150}. 
        (a) The local update rule for Rule 150. $\{p,q,r\}$ denotes the values of the sites in the neighbourhood of $q$, which determines the value of the site $s$ in the next time step.
        (b) The evolution from a single initial down site under Rule 150.
        (c) One of the stable cycles of Rule 150.}
     \label{fig:r150}
\end{figure*}

For 1D elementary CA, the neighbourhood of each lattice site is composed of itself and its two neighbouring sites. 
As a result, in a 3-site neighbourhood and for the finite field $\mathbb{F}_2$, there are 8 possibilities for the states, which give $2^8 = 256$ 1D elementary CA in total \cite{wolfram1983statistical,martin1984algebraic}. For example, Rules 54 and 150 have the update rules 
\footnote{
    The CA we study here are slightly different from those with ``Floquet'' dynamics which have received much attention recently, such as Rule 54~\cite{prosen2016integrability,inoue2018two-extensions,prosen2017exact,gopalakrishnan2018hydrodynamics,buca2019exact,klobas2020space-like,klobas2019time-dependent,alba2019operator,klobas2020matrix}, Rule 150~\cite{gopalakrishnan2018facilitated,gombor2021integrable,wilkinson2022exact-solution} and Rule 201~\cite{wilkinson2020exact,iadecola2020nonergodic}.
    The CA we consider have the traditional synchronous dynamics where all sites are updated simultaneously at every time step, while in Floquet-CA one applies successive 
    partial timesteps of commuting transitions in a ``brickwork'' circuit arrangement. 
}
\begin{equation}
\begin{aligned}
    s = &f_{54}(p, q, r) \;\,= p + q + r + pr \mod{2} \\
    s = &f_{150}(p, q, r) = p + q + r \;\;\;\;\;\;\;\;\; \mod{2} 
\end{aligned}
\end{equation}
where $\{p,q,r\}$ describe the values of the sites in the neighbourhood of the site being updated, see Fig.~\ref{fig:r150}(a). Rule 150 constitutes an example of a linear CA, while Rule 54 is nonlinear. Example trajectories from one initial seed and for a stable cycle are shown in Figs.~\ref{fig:r150}(b, c), respectively.

The periodic structure of a given configuration of linear size $L$ for the time evolution under a certain rule involves a sequence of configurations which get repeated after applying the update rule $M$ times, defining a cycle of period $M$.
The time evolution of linear CA can be described in an algebraic-theoretic way, so that a brute force calculation for the period detection 
is not necessary \cite{martin1984algebraic,jen1988cylindrical,stevens1993transient,stevens1999on-the-construction}. 
For an overview of this method, see Ref.~\cite{sfairopoulos2023boundary} and references therein.
We here present an example of this approach for Rule 150. The local update rule for a row of $L$ sites can be expressed in matrix form as
\begin{equation}
    A_{150} = \begin{bmatrix}
        1 & 1 & 0 & \ldots & 0 & 0 & 1 \\
        1 & 1 & 1 & \ldots & 0 & 0 & 0 \\
        0 & 1 & 1 & \ldots & 0 & 0 & 0 \\
        \vdots & \vdots & \vdots & \ddots & \vdots & \vdots & \vdots \\
        0 & 0 & 0 & \ldots & 1 & 1 & 0 \\
        0 & 0 & 0 & \ldots & 1 & 1 & 1\\
        1 & 0 & 0 & \ldots & 0 & 1 & 1 \\
    \end{bmatrix}.
\end{equation}    
The  matrix $A_{150}$ can be expressed as 
\begin{equation}
    A_{150} = I + S_l + S_r
\end{equation}
with $S_l$ and $S_r$ the left and right shift map, respectively, and where PBC are assumed \cite{calkin2005a-characterization}.
The construction of the ``minimal polynomial'' follows. The order 
of the irreducible polynomials in its decomposition will give the cycle lengths for the cellular automaton. We give the detailed structure of the periods 
of Rule 150 in Table~\ref{tab:periods_150}
for sizes up to $L=40$.

\begin{table}
    \begin{tabular}{cccccccccccc} 
        \toprule
            $L$        & $M$  &&&&&&&&
            $L$        & $M$  \\
        \midrule
            3        & $1$ &&&&&&&&
            22       & {$1, 31, 62$} \\
            4        & {$1, 2$} &&&&&&&&
            23       & {$1, 2047$} \\
            5        & {$1, 3$} &&&&&&&&
            24       & {$1, 2, 4$} \\
            6        & 1 &&&&&&&&
            25       & {$1, 3, 1023$} \\
            7        & {$1, 7 $} &&&&&&&&
            26       & {$1, 21, 42$} \\
            8        & {$1, 2, 4$} &&&&&&&&
            27       & {$1, 7, 511$} \\
            9        & {$1, 7 $} &&&&&&&&
            28       & {$1, 2, 7, 14, 28$} \\
            10       & {$1, 3, 6$} &&&&&&&&
            29       & {$1, 16383$}  \\
            11       & {$1, 31$} &&&&&&&&
            30       & {$1, 3, 5, 6, 10, 15, 30$} \\
            12       & {$1, 2 $} &&&&&&&&
            31       & {$1, 31$} \\ 
            13       & {$1, 21 $} &&&&&&&&
            32       & {$1, 2, 4, 8, 16$} \\
            14       & {$1, 7, 14 $} &&&&&&&&
            33       & {$1, 31$} \\
            15       & {$1, 3, 5, 15$} &&&&&&&&
            34       & {$1, 15, 30$} \\
            16       & {$1, 2, 4, 8 $} &&&&&&&&
            35       & {$1, 3, 7, 21, 4095$} \\
            17       & {$1, 15 $} &&&&&&&&
            36       & {$1, 2, 7, 14, 28$} \\
            18       & {$1, 7, 14$} &&&&&&&&
            37       & {$1, 29127$} \\
            19       & {$1, 511$} &&&&&&&&
            38       & {$1, 511, 1022$} \\
            20       & {$1, 2, 3, 6, 12$} &&&&&&&&
            39       & {$1, 21, 4095$} \\
            21       & {$1, 7, 63$} &&&&&&&&
            40       & {$1, 3, 4, 6, 12, 24$}\\
        \bottomrule
    \end{tabular}
    \caption{
        {\bf Invariant cycles of Rule 150}. 
        Distinct periods of length $M$ of the invariant cycles for systems of linear size $L$ and PBC. Note that this table lists only the distinct periods (including invariant states of period $M=1$), but not their multiplicities. 
    }
    \label{tab:periods_150}
\end{table}

\subsection{2D cellular automata: SPyM-CA}{\label{spymCA}}

The basic theory of CA was first formalized for 1D CA, but extensions to higher dimensions \cite{packard1985two-dimensional} describe 
many fascinating systems, with Conway's Game of Life being a prime example \cite{wolfram1983statistical}. For 2D CA the sites form in general a rectangular lattice. 
In most cases a von Neumann or a Moore neighbourhood are used for the update rule of the given CA \cite{packard1985two-dimensional}.  For a Moore neighbourhood, updating each site involves taking into account the $3 \times 3$ square including both the nearest and the next-nearest neighbours \cite{wolfram2002a-new-kind}, while for a von Neumann neighbourhood only the nearest neighbours are taken into account.

Out of the vast set of possible 2D CA, we will focus on one which, as shown below, relates to the 3D square pyramid model (SPyM) of Refs.~\cite{turner2015overlap,jack2016phase}. The 3D SPyM is a generalization of the 2D triangular plaquette model (TPM) \cite{newman1999glassy,garrahan2001glassy,garrahan2002glassiness} whose associated 1D CA is Rule 60, 
\begin{equation}
    s = f_{60}(p, q, r) = p + q \mod{2}.
\end{equation}
Rule 60 generalizes to 2D to the update rule for the ``SPyM'' CA  
\begin{equation}
    t = f_{\rm SPyM}(p_1, \dots, p_9) = 
    p_{1,1} + p_{1,2} + p_{2,1} + p_{2,2} \mod{2}.
\end{equation}
From now on, the labelling convention $(p_{1,1}, p_{1,2}, p_{2,1}, p_{2,2}) = (p, q, r, s)$ (see Fig.~\ref{fig:SPyM}) will be used.

With the rule defined, the next task concerns the classification of the attractor structure of the model. Since the SPyM rule is linear, its time evolution can be described in an algebraic-theoretic way, 
following Refs.~\cite{sfairopoulos2023boundary,martin1984algebraic,jen1988cylindrical,stevens1993transient,stevens1999on-the-construction,roy-chowdhury1993characterization}. For example, for a CA of size $K \times L = 3\times 3$ we can express the evolution matrix as 
\begin{equation}
    A_{\text{SPyM}} = \begin{bmatrix}
        D & 0 & D \\
        D & D & 0 \\
        0 & D & D \\
    \end{bmatrix},
\end{equation}
with 
\begin{equation}
    D \equiv D_{3\times 3} = \begin{bmatrix}
        1 & 0 & 1 \\
        1 & 1 & 0 \\
        0 & 1 & 1 \\
    \end{bmatrix}.
\end{equation}
This generalizes so that the matrix $A_{\text{SPyM}}$ can be expressed for square systems of odd size as \footnote{We thank Jyrki Lahtonen for this observation.}
\begin{equation}
    A_{\text{SPyM}} =D \otimes D, 
\end{equation} 
where $D$ is the matrix for the evolution of Rule 60.

Computing the periods of the stable cycles for the 2D SPyM-CA is a generalization of the calculation for Rule 60. For example, for lattices where one dimension is a power of two there is a single fixed point (``cycle'' of period 1) with all sites up. Table \ref{tab:periods_SPyM} gives an indicative structure for the periodic behaviour of the SPyM-CA for square initial arrays with $K = L$ 
(there are many more non-square arrangements which we do not discuss here).
The number of ground states for each period can also be verified through the implementation of Floyd's ``\textit{tortoise and hare}'' algorithm \cite{loehr2022the-tortise}.

It is important mentioning that, although SPyM originates from Rule 60, its fixed point structure is different from that of Rule 60; when $K$ or $L$ are a power of 2, we still observe a single fixed point, the trivial one. For other system sizes, however, there might exist multiple fixed points, e.g. for a system size $3\times 3 \times 4$ there exist 4.
The multiplicity of periodic orbits of a given period grows at most subextensively. For example, for a $6\times 6$ lattice there are 10840 periods of length 6, 80 periods of length 3, 120 periods of length 2, and 16 fixed points.

\begin{table}
    \begin{tabular}{cccccccccccc} 
        \toprule
        $L$               & $M$  &&&  \\ 
        \midrule
         3 &   {$1, 3$}  \\
         4 &    {$1$} \\  
         5 &    {$1, 3, 5, 15$} \\
         6 &    {$1, 2, 3, 6$} \\
         7 &    {$1, 7$} \\ 
         8 &    {$1$} \\
         9 &    {$1, 3, 7, 21, 63$} \\
         10 &    {$1, 3, 5, 6, 10, 15, 30$} \\
         11 &    {$1, 31, 341$}  \\
         12 &    {$1, 3, 4, 6, 12$} \\
         13 &    {$1, 63, 91, 273, 819$} \\
         14 &    {$1, 2, 7, 14$} \\
         15 &    {$1, 3, 5, 15$} \\
         16 &    {$1$} \\
         17 &    {$1, 5, 15, 17, 51, 85, 255$} \\
         18 &    {$1, 2, 3, 6, 7, 14, 21, 42, 63, 126$} \\
         19 &    {$1, 511, 9709$} \\
         20 &    {$1, 3, 5, 12, 15, 20, 30, 60$} \\
         21 &    {$1, 3, 7, 9, 21, 63$} \\
         22 &    {$1, 31, 62, 341, 682$} \\
         23 &    {$1, 89, 2047$} \\
         24 &    {$1, 3, 6, 8, 12, 24$} \\
         25 &    {$1, 3, 5, 15, 775, 1023, 2325, 5115, 8525, 25575$} \\
         26 &    {$1, 63, 91, 126, 182, 273, 546, 819, 1638$} \\
         27 &    {$1, 3, 7, 21, 63, 511, 1533, 1971, 4599, 13797$} \\
         28 &    {$1, 4, 7, 14, 28$} \\
         29 &    {$1, 3683, 16383, 158369, 475107$} \\
         30 &    {$1, 2, 3, 5, 6, 10, 15, 30$} \\
         \bottomrule
    \end{tabular}
    \caption{
        {\bf Invariant cycles of the SPyM-CA}.
        Distinct periods of length ${\cal M}$ for systems of size $ \times L$ for $K = L$ and PBC. 
    }
    \label{tab:periods_SPyM}
\end{table}
\section{Classical spin models from CA}{\label{Section_classicalmodels}}

\subsection{Two-dimensional spin models}

We first construct the 2D spin models on the square lattice whose ground states are given by 1D CA rules. This method generalizes Ref.~\cite{sfairopoulos2023boundary} to all 1D elementary CA.
Based on these CA we construct the ``simplest'' (i.e.\ most local and least number of interaction terms) classical Hamiltonians whose minimum energy configurations are given by the stable cycles of the given CA. In Appx.~\ref{appendix:all-models}  we list these classical Hamiltonians for the 256 elementary CA.

The 2D spin models we consider live on a square lattice, where the time direction of the CA maps into the second spatial dimension of the spin model. The interactions in the spin Hamiltonian are between sites that form the CA neighbourhood, cf.\ Fig.~\ref{fig:r150}a. This allows for up to four-spin interactions, including triangular plaquette interactions (if we shear a triangular lattice into a square one). The classification of all the 256 spin models given in Appx.~\ref{appendix:all-models} is obtained from the combination of eight ``fundamental'' models. These models are constructed based on the eight possible elementary CA with linear update rules. These are
\begin{align}
    &E_0  \;\;\;\;= - \sum_{s} \sigma_s 
    \label{m0}    
    \\
    &E_{240}\;   = - \sum_{\{p, s\} \in \Isingdiagonal}       \sigma_p \sigma_s     \label{m240}    
    \\
    &E_{204} \;  = - \sum_{\{q, s\} \in \Isingvertical}       \sigma_q \sigma_s     \label{m204}  
    \\
    &E_{170}  = - \sum_{\{r, s\} \in \Isingdiagonalsecond}       \sigma_r \sigma_s     \label{m170}   
    \\
    &E_{60} \, \, \;= - \sum_{\{p, q, s\} \in \trigonright}    \sigma_p \sigma_q \sigma_s    \label{m60}    
    \\
    &E_{90} \; \, \,= - \sum_{\{p, r, s\} \in \trigonmiddle}    \sigma_p \sigma_r \sigma_s     \label{m90}    
    \\
    &E_{102} \;     = - \sum_{\{q, r, s\} \in \trigonleft}    \sigma_q \sigma_r \sigma_s     \label{m102}    
    \\
    &E_{150} \;     = - \sum_{\{p, q, r, s\} \in \tetrapleuro} \sigma_p \sigma_q \sigma_r \sigma_s,
    \label{m150}    
\end{align} 
where $\sigma_i = 1-2x_i = \pm 1$ with $x_i = \{p,q,r,s\}$ indicating both the location of the spin degrees of freedom but also the state of the CA site (up/down being 0/1). The subscript in the Hamiltonian above indicates the associated CA rule. We see from the above that Rule 0 corresponds to a non-interacting system; Rules 240, 204 and 170 to 1D Ising models along the diagonal or vertical dimension; Rules 60 and 102 to the TPM (and its spatial reflection); Rule 90 to the three-spin interaction where the spin with index $q$ does not participate in the interaction; and Rule 150 to a four-spin interacting model, mapping to the product of Rule 90 with the spin labelled $q$. All the other 2D models related to the elementary CA
are obtained via linear combinations of the models in Eqs.~\eqref{m0}-\eqref{m150}, see Appx.~\ref{appendix:all-models}.

The minimum energy configurations of the models defined by Eqs.~\eqref{m0}-\eqref{m150} are found by using the corresponding CA. The exact set of ground states depends on the nature of the boundary conditions of the spin model. The one we study is for
PBC in both lattice directions. In this case the ground states are given by the limit cycles of the associated CA of size $L$ with PBC in space, cf.\ Table~\ref{tab:periods_150}.
Similarly, for spin models with open boundary conditions (OBC) in the horizontal (or $x$) dimension, but PBC in the vertical (or $y$) direction, the ground states come from the commensurate limit cycles of the 
corresponding CA with OBC in space. Similarly, PBC in the space dimension of the CA and OBC in its time direction would imply that all initial configurations are accepted as ground states for the related spin model, as no further criterion needs to be applied. The number of ground states in this case is $2^L$. 
For a lattice with OBC in both dimensions, the number of classical ground states is $2^{L + M - 1}$ or $2^{L + 2M - 2}$ depending on the span of the interaction, due to the freedom in choosing the first row and the first one or two columns of sites of the $L\times M$ lattice. Here, we do not consider antiperiodic boundary conditions or any other choice for the boundaries.

Although our focus was on spin models constructed from elementary CA, there is always the possibility of generalizing the above results. We give in Appx.~\ref{appendix:other-CA} a very brief overview of some classical spin models which extend the notion of elementary CA to larger neighbourhoods or higher dimensions in order to show the versatility of our approach.

\begin{figure}
   \centering
   \includegraphics[width=0.5\columnwidth]{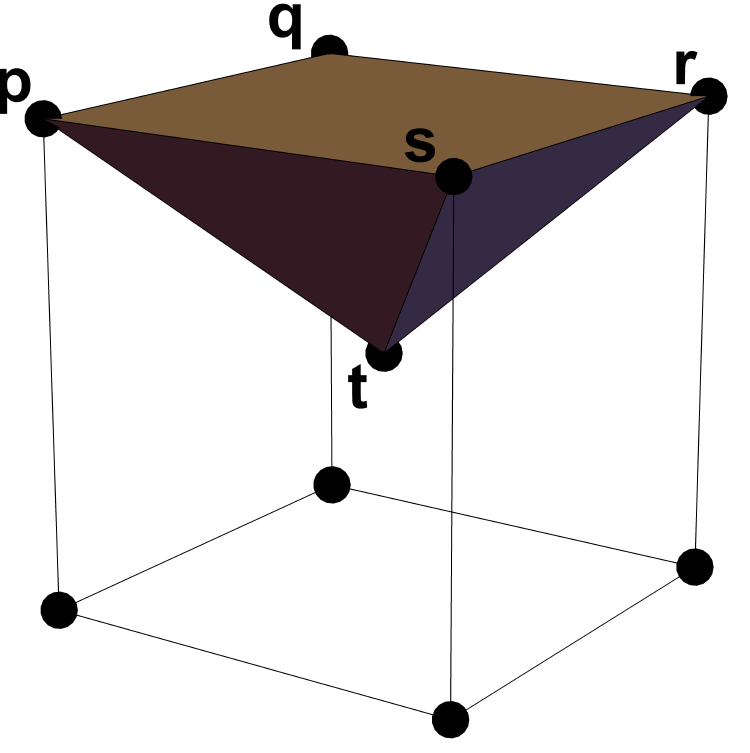}
   \caption{
      {\bf Square Pyramid Model.}
      5-spin interaction in the SPyM. 
   }
   \label{fig:SPyM} 
\end{figure}

\subsection{3D spin models: SPyM}

For the connection between 2D CA and 3D spin models we focus on the classical model introduced in Ref.~\cite{heringa1989phase}, termed the square pyramid model (SPyM) in Refs.~\cite{turner2015overlap,jack2016phase}. Initially, it was studied as a model of glasses that generalizes the TPM to three dimensions. The classical Hamiltonian of the SPyM is 


\tdplotsetmaincoords{70}{120}
\noindent
\begin{equation}
    \opcatSPyM{E} = - J \sum_{{\textstyle\mathstrut} \{p, q, r, s, t\} \in \raisebox{-0.3ex}{\begin{tikzpicture}[tdplot_main_coords,line cap=butt,line join=round,c/.style={circle,fill,inner sep=1pt},
            declare function={a=1.0/6;h=1.4/5;}]
            \path
            (0,0,0) coordinate (A)
            (a*2.3/2,-1/20,0) coordinate (B)
            (a,1.7*a/2,0) coordinate (C)
            (-0.0005,a*2.5/2,0) coordinate (D)
            (0,0,-h)  coordinate (S);
            \draw (S) -- (D) -- (C) -- (B) -- cycle (S) -- (C);
            \draw (B) -- (A) --(D);
            \draw[densely dotted] (A) -- (S);
        \end{tikzpicture}}}
        \sigma_p \sigma_q \sigma_r \sigma_s \sigma_t,
    \label{ESPyM}
\end{equation}
where the spins interact on downward pointing pyramids on a BCC lattice, see Fig.~\ref{fig:SPyM}.

The Hamiltonian Eq.~\eqref{ESPyM} is the simplest ``parent'' Hamiltonian whose classical ground states are determined from the dynamics of the 2D SPyM-CA of Sec.~\ref{spymCA}. The same considerations relating to boundary conditions as for 2D spin systems apply here. For the SPyM with PBC, the minima are given by the fixed points and commensurate invariant cycles of the SPyM-CA, see Table~\ref{tab:periods_SPyM}. For the SPyM with OBC in a BCC lattice of size $K \times L \times M$, the number of ground 
states is $2^{K+L+M-2}$. A similar analysis based on 2D CA can be performed for the ground states of the other 3D spin models of Ref.~\cite{heringa1989phase}.

\subsection{Baxter-Wu model and CA Rule 60}{\label{BWclassical}}

\begin{figure}
    \includegraphics[width=0.7\columnwidth]{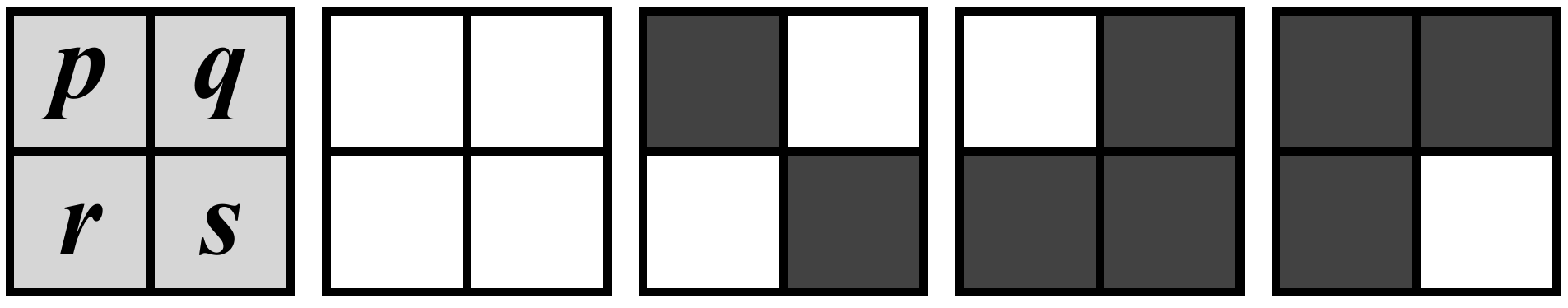}
    \caption{
      {\bf Baxter-Wu model.} The labelling for the $2\times2$ blocks, and the accepted ground state blocks, where white/black indicate up/down sites.
      }
    \label{fig:BW}
\end{figure}

The Baxter-Wu (BW) model \cite{baxter1973exact,baxter1974ising,capponi2014baxter-wu,baxter2016exactly} is a spin model on the triangular lattice with three-body interactions between both upward and downward pointing triangular plaquettes. Its Hamiltonian reads
\begin{equation}
   E_{\rm BW} = -\sum_{\hspace{0pt minus 1fil} \downtriangle} \sigma_p \sigma_q \sigma_s -\sum_{\hspace{0pt minus 1fil} \uppertriangle} \sigma_s \sigma_r \sigma_p ,
   \label{bw}
\end{equation}
where the location of the spins is sketched in Fig.~\ref{fig:BW} (where again we have represented the triangular lattices sheared into a square lattice). The energy function above is the sum of a TPM (downward-pointing plaquettes) and a vertically inverted TPM (upward-pointing plaquettes). While the ground states of the TPM can be inferred from the 1D CA Rule 60, the minimum energy configurations of Eq.~\eqref{bw} are given by {\em two counter-propagating} Rule 60 CA, see Fig.~\ref{fig:BW}: the forward propagating Rule 60 minimizes the first sum in Eq.~\eqref{bw}, while the backward propagating minimizes the second sum. The minimization of Eq.~\eqref{bw} involves the simultaneous minimization of both CA trajectories, which imposes the restriction 
\begin{equation}
   p + q = s , \;\;\;
   s + r = p \mod{2},
\end{equation}
where as before $\{p,q,r,s\}$ represent the values of the CA sites, see Fig.~\ref{fig:BW}.  

Based on the above considerations, the BW model has always the trivial (all spins up) ground state regardless of boundary conditions. For PBC, a BW model of size $L \times M$ has four ground states if $L$ and $M$ are multiples of $3$ and only the trivial ground state otherwise. For OBC the BW model has always 4 ground states.

\begin{figure*}
   \centering
   \begin{subfigure}[b]{0.32\textwidth}
      \includegraphics[width=\textwidth]{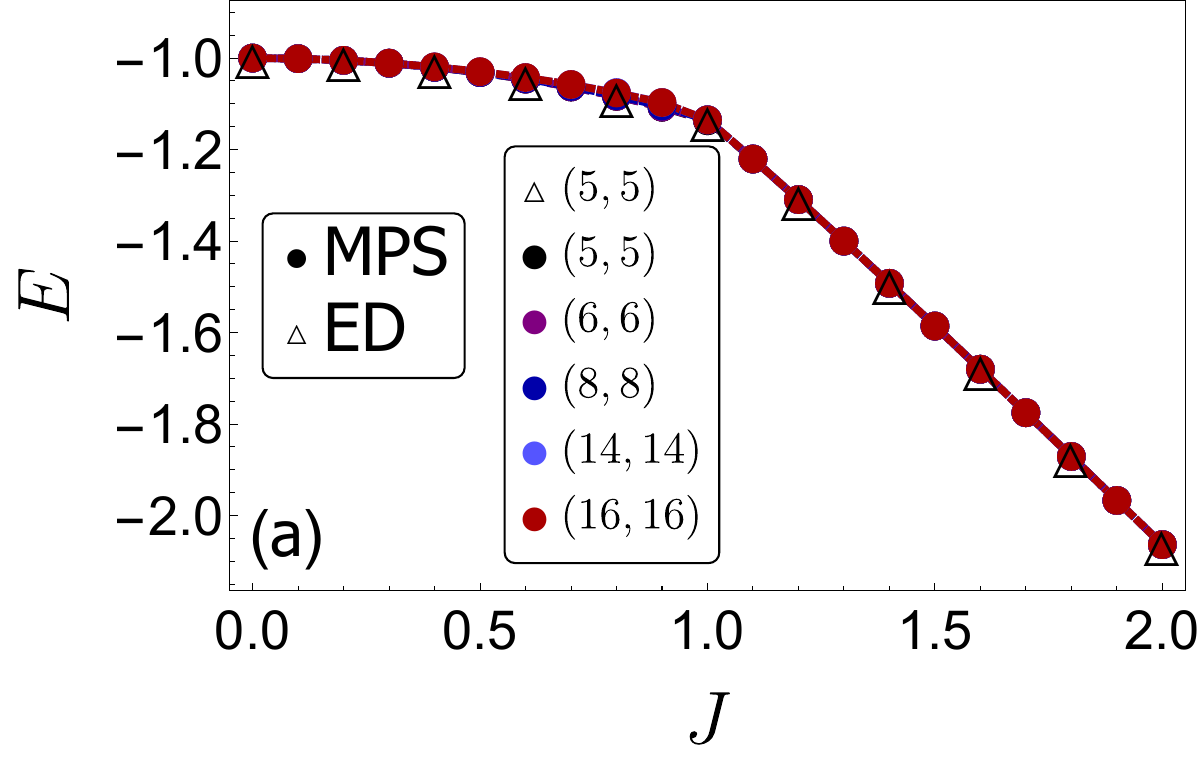}
   \end{subfigure}
   \begin{subfigure}[b]{0.32\textwidth}
      \includegraphics[width=\textwidth]{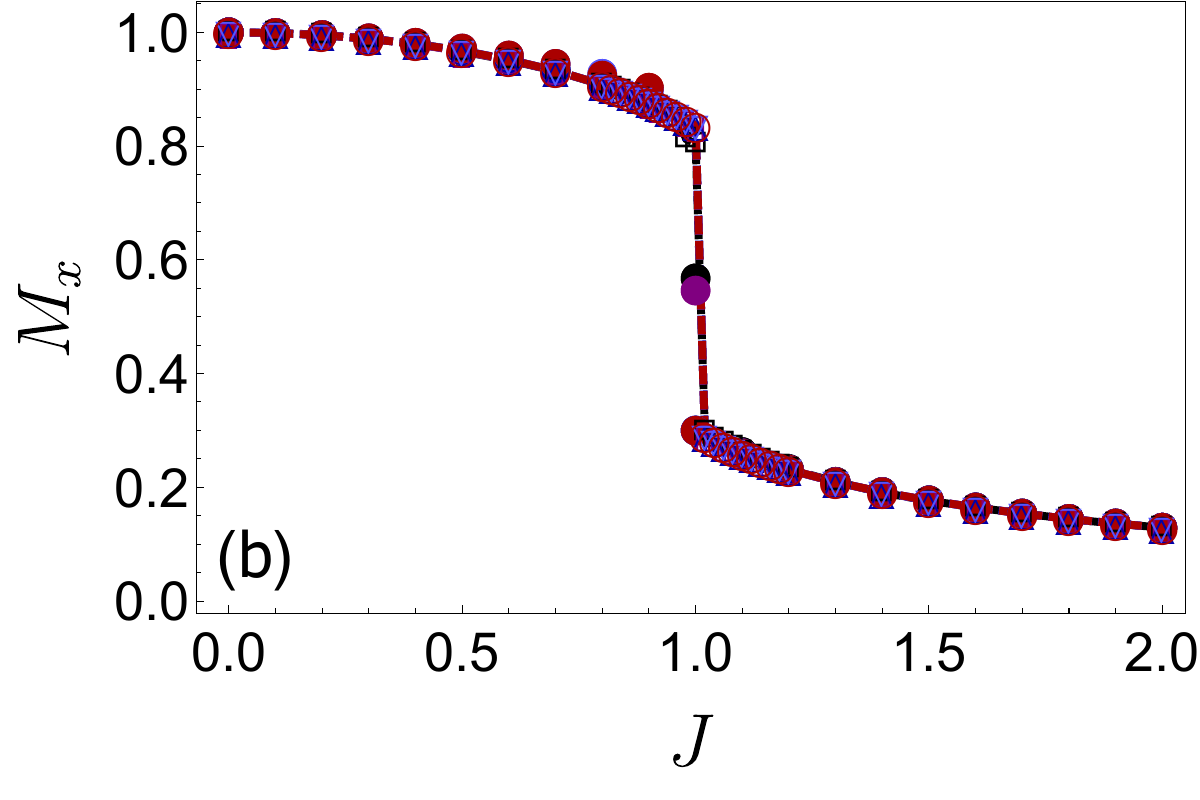}
   \end{subfigure}
   \begin{subfigure}[b]{0.32\textwidth}
      \includegraphics[width=\textwidth]{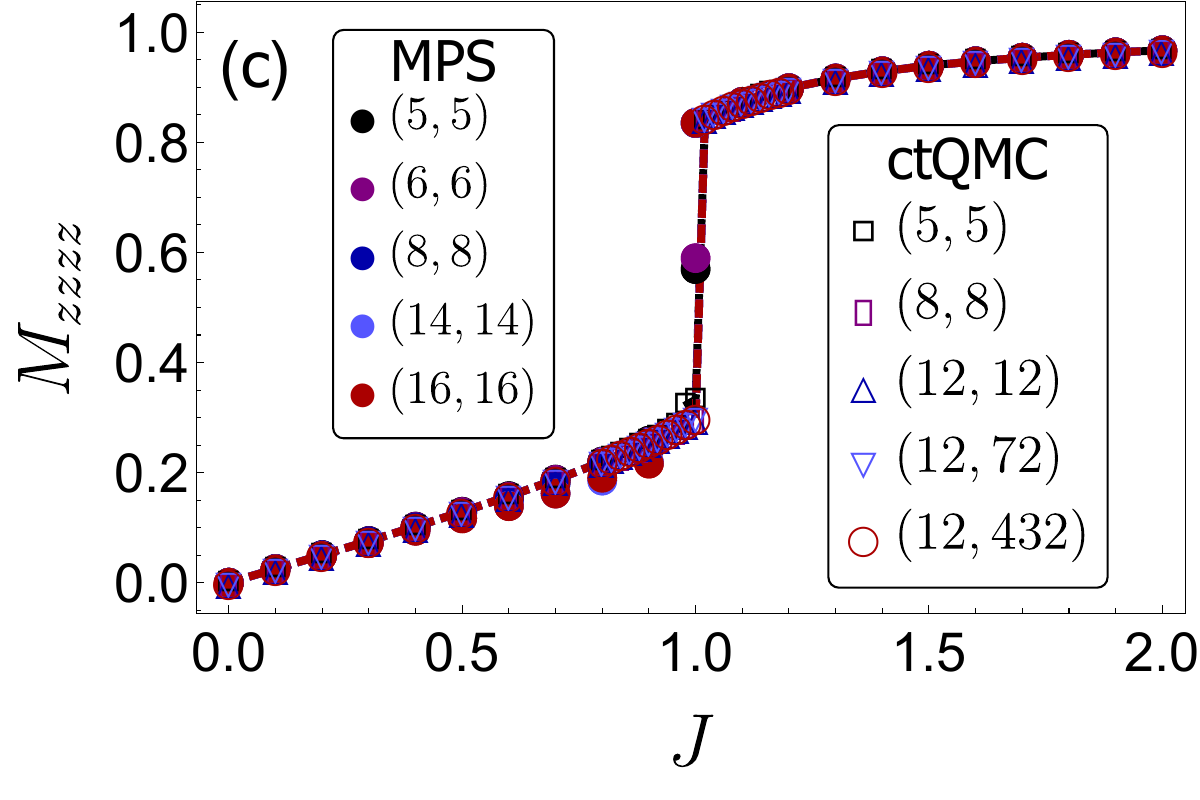}
   \end{subfigure}
   \begin{subfigure}[b]{0.32\textwidth}
      \includegraphics[width=\textwidth]{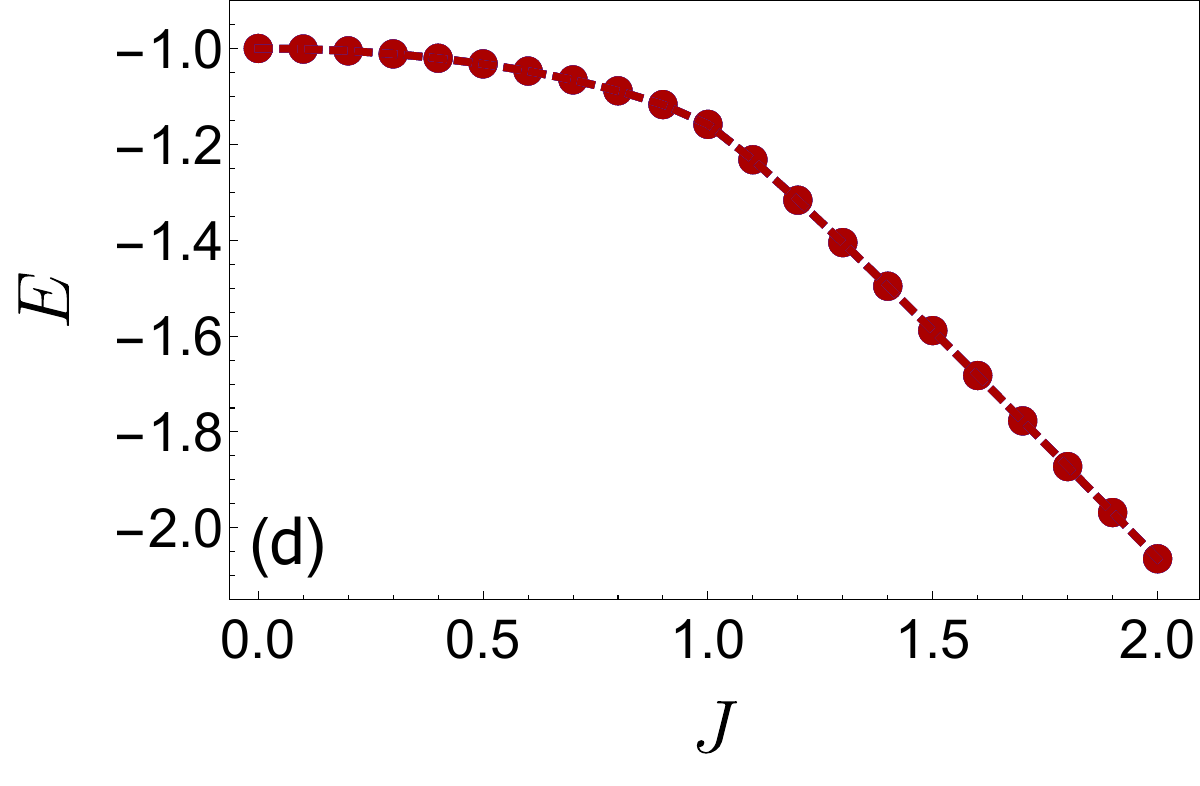}
   \end{subfigure}
   \begin{subfigure}[b]{0.32\textwidth}
      \includegraphics[width=\textwidth]{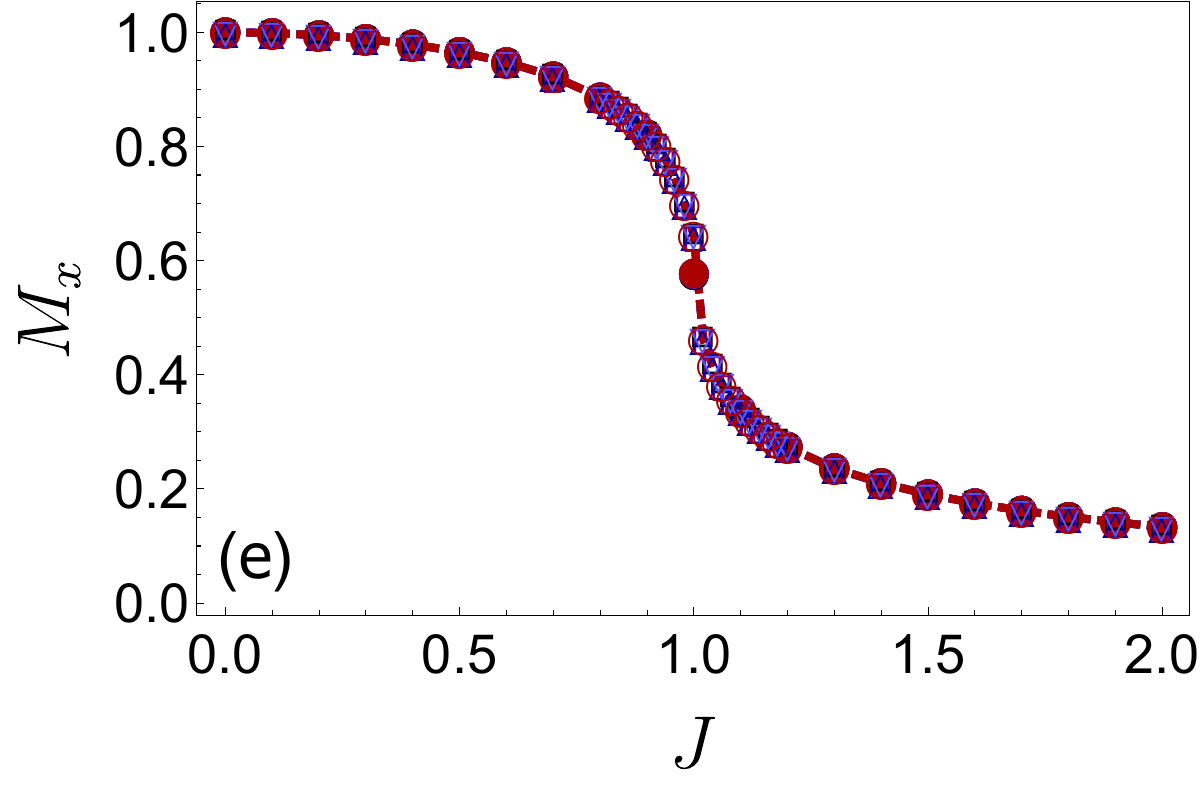}
   \end{subfigure}
   \begin{subfigure}[b]{0.32\textwidth}
      \includegraphics[width=\textwidth]{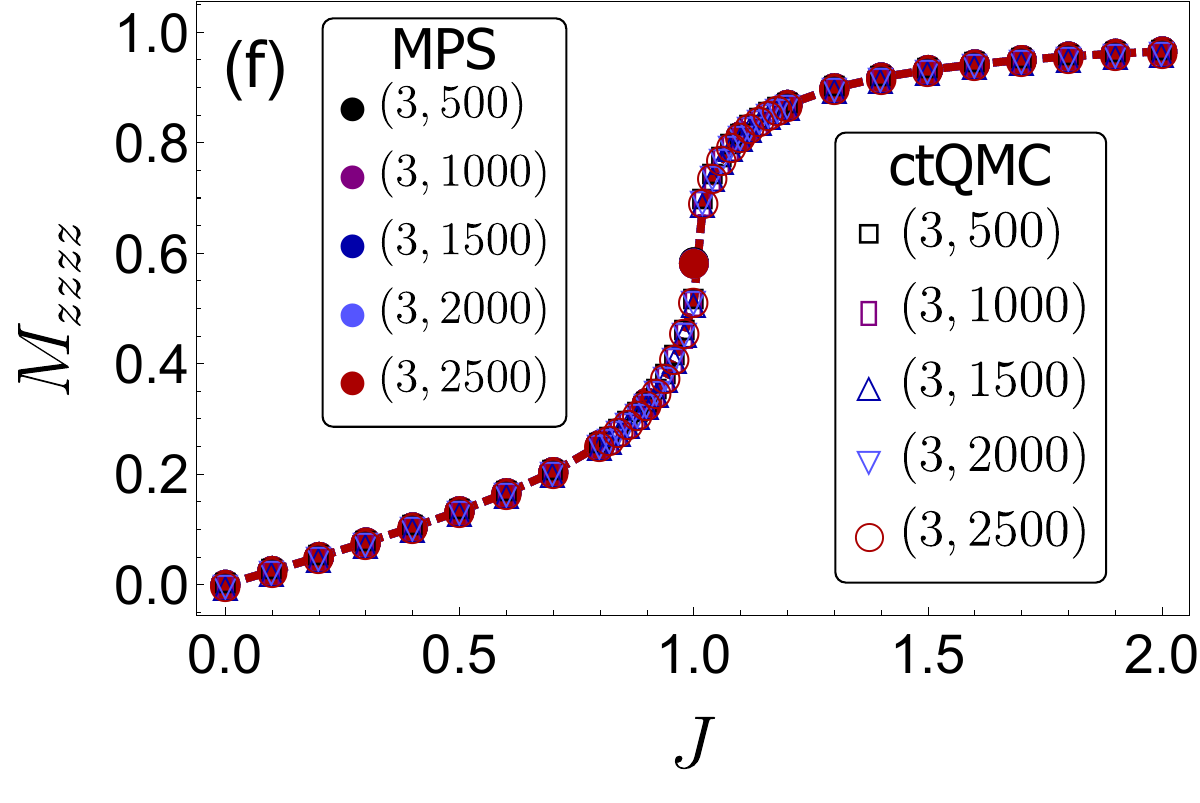}
   \end{subfigure}
   \caption{
      {\bf Quantum phase transition of $H_{150}$ for PBC.} 
      (a) The normalized ground state energy $E$ as a function of $J$ for square systems, $L \times L$. Empty symbols are from ED while filled symbols are from numerical MPS. (b) The transverse magnetization $M_x$ as a function of $J$. Here open symbols are from ctQMC simulations, and include wide rectangular systems, $L \times M$. (c) Same for the average four-spin interaction $M_{zzzz}$ (see legends in panel).
      (d-f) Same but for thin strip geometries, as indicated in panel (f). 
      }
   \label{fig:q150}
\end{figure*}

\section{Quantum spin models and ground state phase transitions}{\label{Quantum models}}

In this section, we minimally couple the models of Sec.~\ref{Section_classicalmodels} by adding a transverse field to their Hamiltonians, 
\begin{equation}
   H_\mu = J E_\mu(Z) - h \sum_{p=1}^N X_p, 
   \label{H} 
\end{equation}
where we use $H_\mu$ to denote the quantum Hamiltonian, and  $E_\mu(Z)$ the classical energy function for the spin system related to CA with rule $\mu$, cf. Eqs.\eqref{m0}-\eqref{bw}, with the Ising spin $\sigma_i$ at site $i$ replaced by the Pauli operator $Z_i$. As expected, this combination of terms creates 
competing orders between the classical energy, which tends to align the spins in the $z$-spin direction of the minimum energy configurations, and the transverse field ones. This competition gives rise to a phase transition in the ground state, controlled by the ratio $J/h$. 

For the models of Eqs.\eqref{m0}-\eqref{bw}, while the configurations which minimize the classical energy do not directly give the quantum ground states in the presence of a field, they do however determine the symmetries of the quantum models \cite{sfairopoulos2023boundary}, and thus we can use the information from the periodic structure of the related CA to infer the existence or not of spontaneous symmetry breaking (SSB). We claim that in most of the spin models the quantum phase transition is of first order in the thermodynamic limit, with the optional addition of SSB for the system size sequences where there are multiple classical ground states. The only models where a continuous quantum phase transition is expected concerns the transverse field Ising model (TFIM) which is a 2-XORSAT instance \cite{mezard2009information}, while here we focus on p-XORSAT instances with $p \geq 3$. This is also the scenario that we found in Ref.~\cite{sfairopoulos2023boundary} for the specific case of the quantum TPM through its connection to Rule 60. For verifying our claims, we present a wealth of results from numerical simulations obtained using exact diagonalisation (ED) \cite{sandvik2010computational} (for systems up to 25 sites), matrix product state (MPS) methods \cite{schollwock2011the-density-matrix,stoudenmire2012studying,fishman2022the-itensor,fishman2022the-itensor2} (with bond dimensions up to 1000), and continuous-time quantum Monte Carlo (ctQMC) simulations \cite{causer2024rejection-free,beard1996simulations,krzakala2008path-integral,mora2012transition} (with $\beta=100$ or bigger).

\subsection{2D quantum spin models}

The quantum spin model from Eq.~\eqref{H} with interaction energies related to Rules 170, 204 and 240, Eqs.~\eqref{m240}-\eqref{m170}, provides multiple copies of the 1D TFIM embedded in a 2D lattice, while for Rule 0 we have a non-interacting spin system with longitudinal and transverse fields. The model connected to Rule 60, Eq.~\eqref{m60}, is the quantum TPM studied in Ref.~\cite{sfairopoulos2023boundary}, and Rule 102, Eq.~\eqref{m102}, is directly related to it by a reflection symmetry. Rule 90 in Eq.~\eqref{m90}, although technically distinct from Rule 60 when PBC are used, it displays very similar behaviour. We therefore choose to study Rule 150, Eq.~\eqref{m150}. 

The Hamiltonian for the quantum spin model connected to CA Rule 150 is
\begin{equation}
   H_{150} = -J \sum_{\{p, q, r, s\} \in \tetrapleuro} Z_p Z_q Z_r Z_s - h \sum_i X_i .
   \label{h150}
\end{equation}
Similar to the case of the quantum Newman-Moore model \cite{vasiloiu2020trajectory}, we can show (following for example Ref.~\cite{xu2004strong-weak,vasiloiu2020trajectory}) that $H_{150}$ has a duality that exchanges the interaction and field terms in Eq.~\eqref{h150} and flips the interaction term. This suggests that, if a single quantum phase transition exists, it will be observed on the self-dual point $J=h$ \cite{cobanera2010unified,cobanera2011the-bond-algebraic,vasiloiu2020trajectory,2020_Tantivasadakarn,seiberg2024majorana,seiberg2024non-invertible}. 
The relevant observables to describe the transition are the normalized transverse magnetization, $$\opcatMx{M} = \frac{1}{N} \sum\limits_i^N X_i,$$ the longitudinal magnetization, $$\opcatMz{M} = \frac{1}{N} \sum\limits_i^N Z_i,$$ 
and the four-spin interaction operator $$\opcatMzzzz{M} = \frac{1}{N} \sum\limits_{\{p, q, r, s\} \in \tetrapleuro} Z_p Z_q Z_r Z_s.$$

Figure~\ref{fig:q150} provides numerical evidence for the quantum phase transition. We use  $h=1.0$ without loss of generality.
Panel (a) shows the ground state energy per unit length as a function of $J$ for square systems $L \times L$ and PBC from ED and numerical MPS, showing a pronounced change in slope around $J=1.0$.
Panels (b) and (c) show the average transverse magnetization, $\opcatMx{M}$, and the average four-spin interaction, $\opcatMzzzz{M}$, for both square and rectangular systems from numerical MPS and ctQMC simulations. 
The jump in these observables close to the $J=1.0$ point provides a clear indication of a first-order transition. Panels (d-f) shows 
the same quantities for quasi-1D geometries, where the signatures of the quantum phase transition are not as clear.
The difference in the appearance of the transition in the thin strip geometries of Fig~\ref{fig:q150}(d-f) to the square or wider rectangular ones of  
panels (a-c) is similar to the one observed for the quantum Newman-Moore model \cite{sfairopoulos2023boundary}. The case for OBC is discussed in Appx.~\ref{appendix:numerics}. 

The nature of the quantum phase transition of $H_{150}$ is related to the low-lying excitations above the ground state, but also the symmetries of the model and their spontaneous breaking (cf.\ Ref.~\cite{sfairopoulos2023boundary}). The symmetries of $H_{150}$ can be obtained as follows: given a system of size $L \times M$, for each limit cycle of Rule 150 commensurate with those dimensions, cf.\ Table~\ref{tab:periods_150}, 
we construct a symmetry operator as the product of $X$-Pauli matrices acting on the sites where the classical ground state (i.e., the CA cycle) differs from the trivial ground state. This is similar to what was done for the quantum TPM, $H_{60}$, in Ref.~\cite{sfairopoulos2023boundary}, and is an approach applicable to all models for Eqs.~\ref{m0}-\ref{m150}. For the case of $H_{150}$, given the quartic interactions, there is also the global spin-flip symmetry. While the lowest energy excitations are difficult to study systematically for arbitrary system sizes, there are clear indications of both the avoided gap crossing that gives rise to the first-order phase transition from small systems and the symmetry breaking, in parallel to Ref.~\cite{sfairopoulos2023boundary}. We present this analysis in Appx.~\ref{appendix:low-lying}.

\begin{figure*}
   \begin{subfigure}[b]{0.24\textwidth}
      \includegraphics[width=\textwidth]{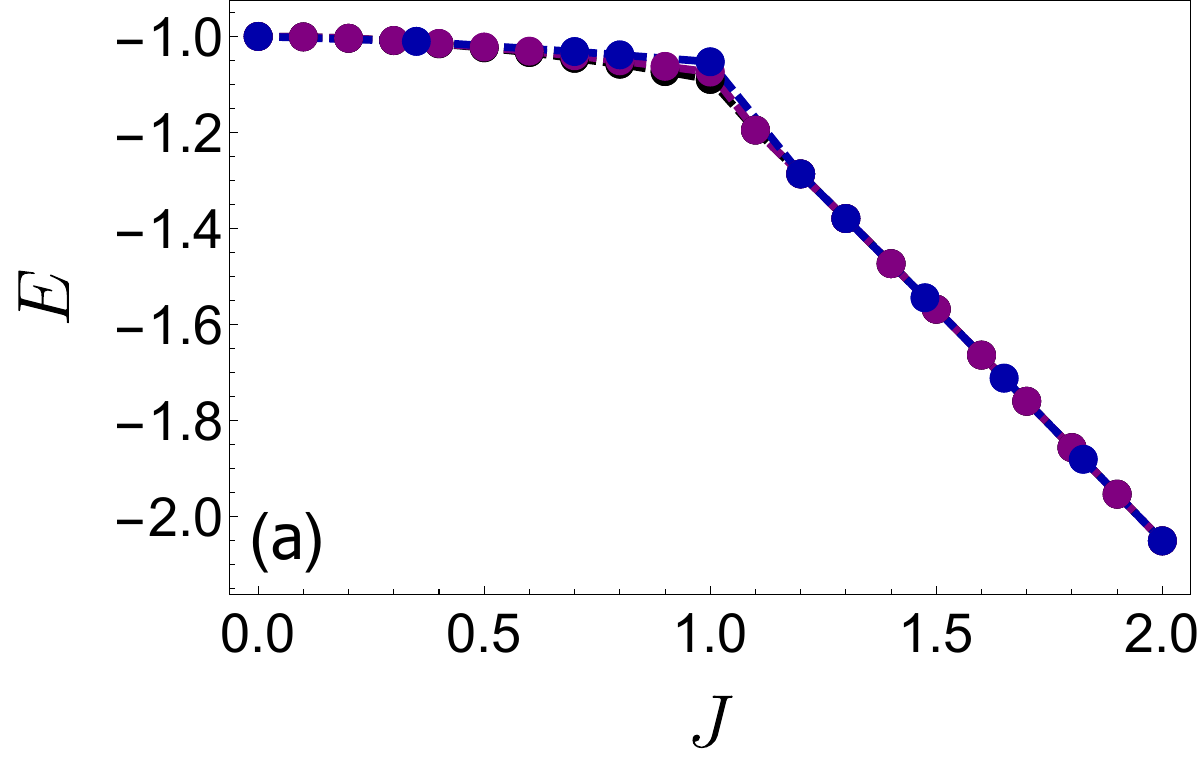}
   \end{subfigure}
   \begin{subfigure}[b]{0.24\textwidth}
      \includegraphics[width=\textwidth]{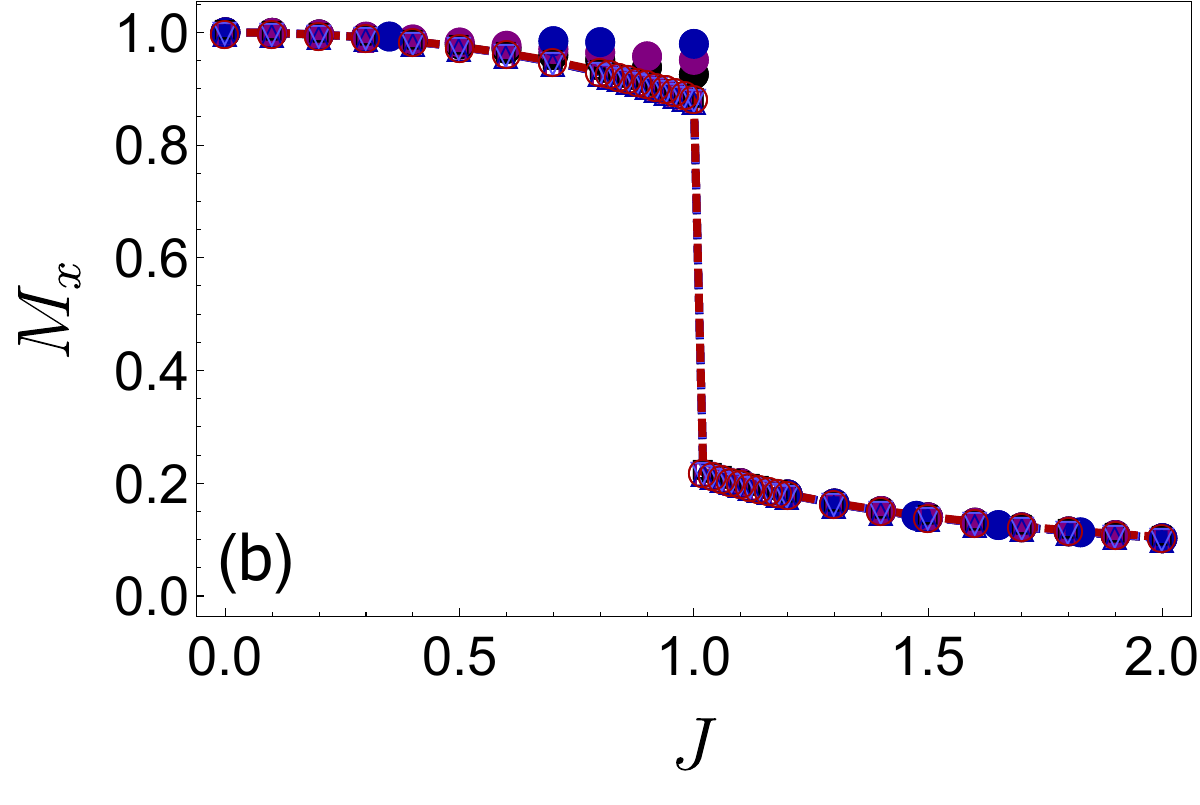}
   \end{subfigure}
   \begin{subfigure}[b]{0.24\textwidth}   
      \includegraphics[width=\textwidth]{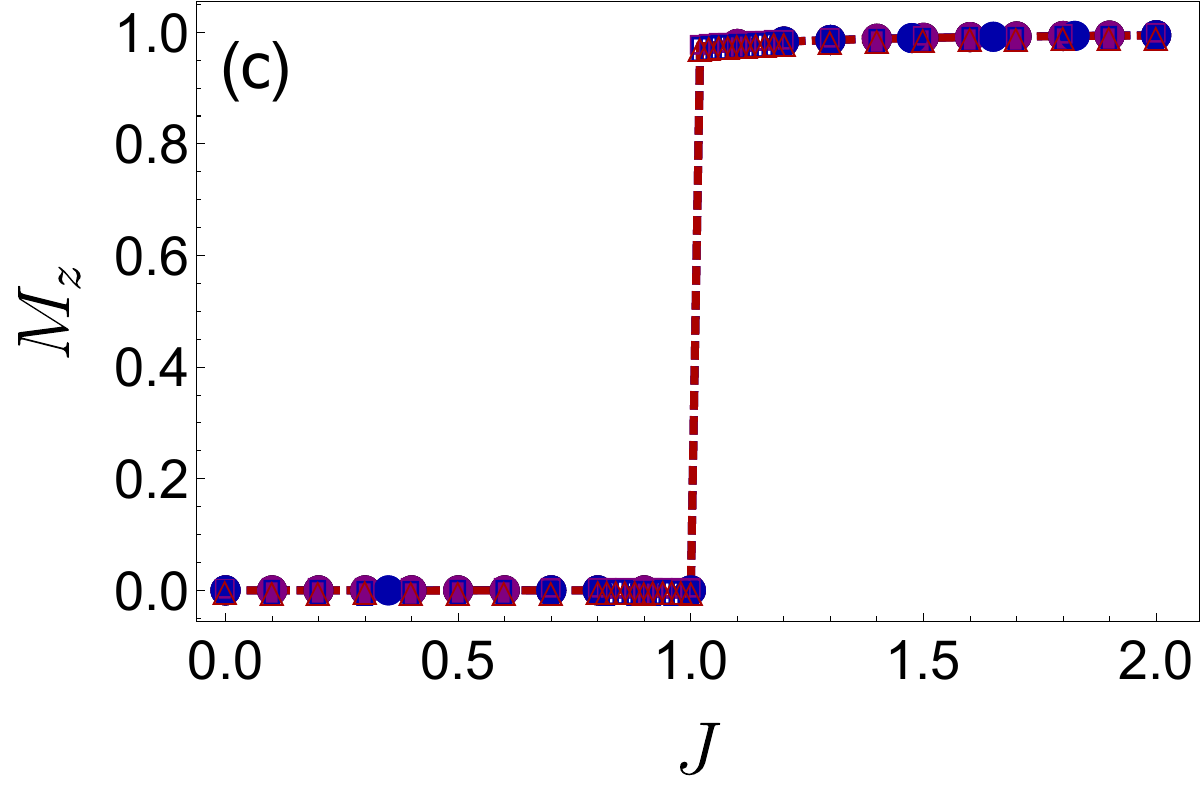}
   \end{subfigure}
   \begin{subfigure}[b]{0.24\textwidth}   
      \includegraphics[width=\textwidth]{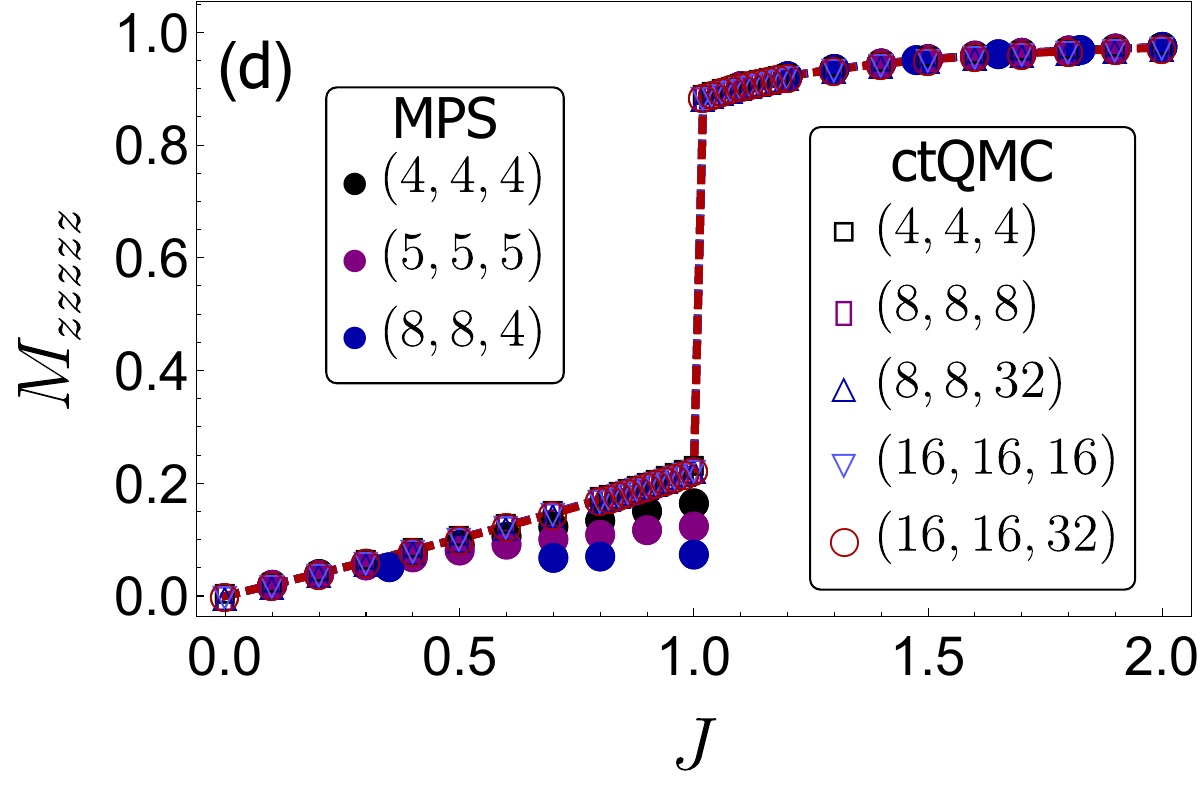}
   \end{subfigure}
   \caption{
      {\bf Quantum phase transition of $\opcatSPyM{H}$ for PBC.}
      (a) The ground state energy per unit size as a function of $J$ for system sizes $L \times L \times M$ (square and rectangular). Filled symbols are from numerical MPS.
      (b) Transverse magnetization $M_x$. Filled symbols are from numerical MPS and empty symbols from ctQMC.
      (c,d) Same for longitudinal magnetization $M_z$, and average five-spin interaction $M_{zzzzz}$, respectively.
      }
    \label{fig:qSPyM}
\end{figure*}

\subsection{3D models: quantum SPyM}

As an example of a 3D quantum spin model we consider the Hamiltonian Eq.~\eqref{H} with the interaction as in Eq.~\eqref{ESPyM}. While the classical SPyM has been considered in the literature, see Refs.~\cite{heringa1989phase,jack2016phase}, to our knowledge, the quantum SPyM has not been studied before. The Hamiltonian for this quantum SPyM reads
\tdplotsetmaincoords{70}{120}

\begin{equation}
    \opcatSPyM{H} = - J \sum_{{\textstyle\mathstrut} \{p, q, r, s, t\} \in \raisebox{-0.3ex}{\begin{tikzpicture}[tdplot_main_coords,line cap=butt,line join=round,c/.style={circle,fill,inner sep=1pt},
      declare function={a=1.0/6;h=1.4/5;}]
      \path
      (0,0,0) coordinate (A)
      (a*2.3/2,-1/20,0) coordinate (B)
      (a,1.7*a/2,0) coordinate (C)
      (-0.0005,a*2.5/2,0) coordinate (D)
      (0,0,-h)  coordinate (S);
      \draw (S) -- (D) -- (C) -- (B) -- cycle (S) -- (C);
      \draw (B) -- (A) --(D);
      \draw[densely dotted] (A) -- (S);
  \end{tikzpicture}}}
  Z_p Z_q Z_r Z_s Z_t - h \sum_{p} X_p,
    \label{HSPyM}
\end{equation}
where the interactions are described in Fig.~\ref{fig:SPyM}. The number of classical ground states is given by the cycles of the 2D SPyM-CA, see Table~\ref{tab:periods_SPyM}, which in turn define the symmetries of Eq.~\eqref{HSPyM}. 

The quantum SPyM is a direct generalization of the quantum Newman-Moore model to 3D. As for the 2D models, it has a duality that exchanges $J$ and $h$, and numerics suggest a quantum phase transition at the self-dual point $J=h$,
see Fig.~\ref{fig:qSPyM}: for all the sizes studied the numerics for PBC indicate a first-order transition at $J=h$ in the large size limit. However, in cases where the system allows multiple classical minima from the cycles of the SPyM-CA, we expect the first-order transition to be accompanied by SSB of the model's symmetries. 
We have verified the numerics of Fig.~\ref{fig:qSPyM} with ED for systems up to 24 spins. While one would expect MPS approximations to rapidly decrease in accuracy in 3D due to the unfavourable scaling of the entanglement with system size \cite{schollwock2011the-density-matrix,stoudenmire2012studying}, Fig.~\ref{fig:SPyM} shows that there is reasonable agreement (for simulations of up to 256 spins) with ctQMC simulations which, in turn, allow one to reach systems of nearly 10000 spins.

\subsection{The Baxter-Wu model in a transverse field}

If in Eq.~\eqref{H} we use Eq.~\eqref{bw} we obtain the quantum Baxter-Wu model
\begin{equation}
   \opcatBW{H} = 
      - J \sum_{\hspace{0pt minus 1fil} \downtriangle} Z_p Z_q Z_s 
      - J \sum_{\hspace{0pt minus 1fil} \uppertriangle} Z_s Z_r Z_p 
      - h \sum_p X_p.
   \label{HBW}
\end{equation}
The Baxter-Wu model in a transverse field was studied in Ref.~\cite{capponi2014baxter-wu} via stochastic series expansions (SSE) for sizes up to $15 \times 15$, finding evidence for a phase transition at $h \approx 2.4$.

\begin{figure*}
   \begin{subfigure}[b]{0.3\textwidth}
       \includegraphics[width=\textwidth]{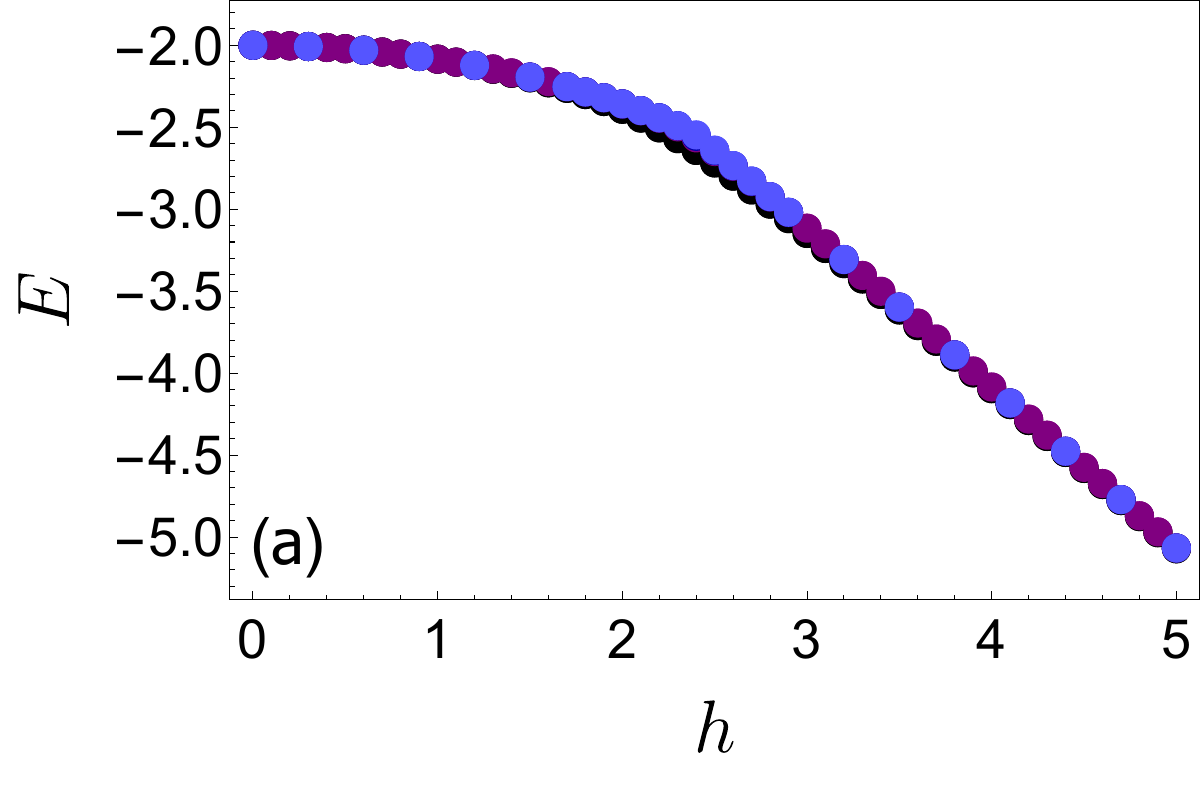}
   \end{subfigure}
    \begin{subfigure}[b]{0.3\textwidth}
      \includegraphics[width=\textwidth]{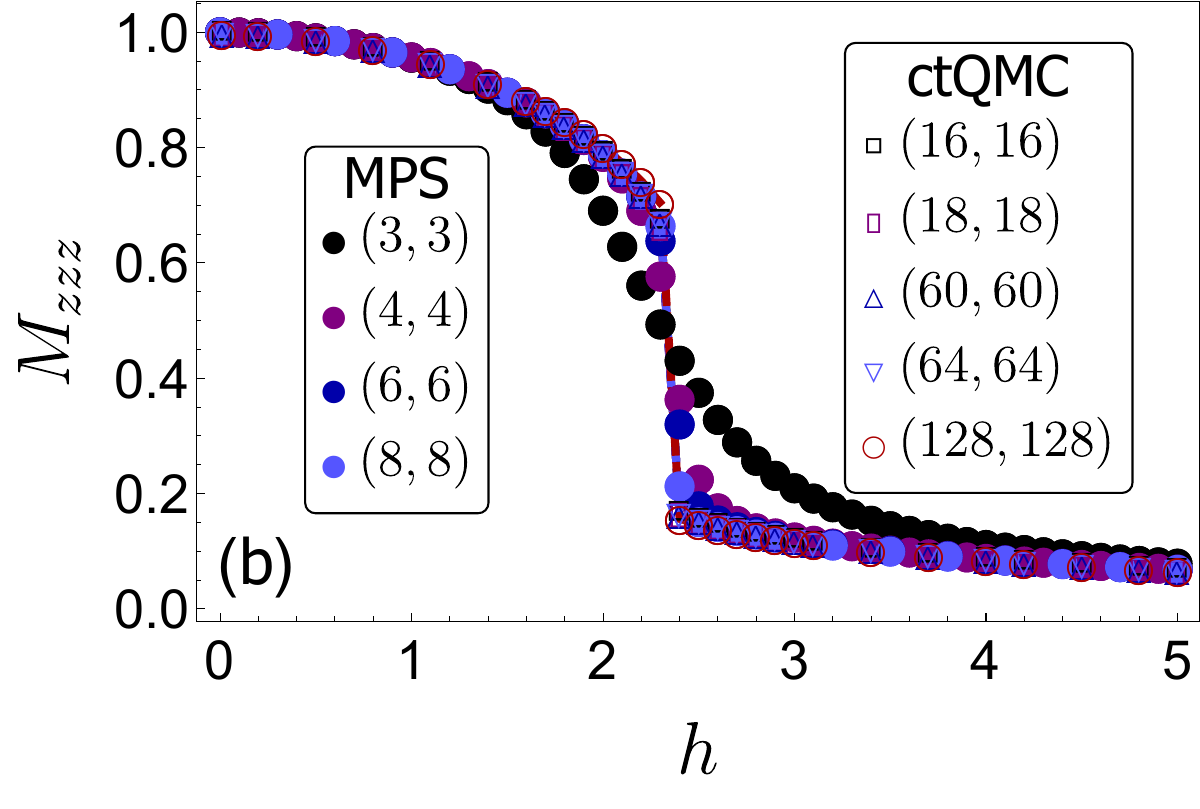}
   \end{subfigure}
   \begin{subfigure}[b]{0.3\textwidth}
      \includegraphics[width=\textwidth]{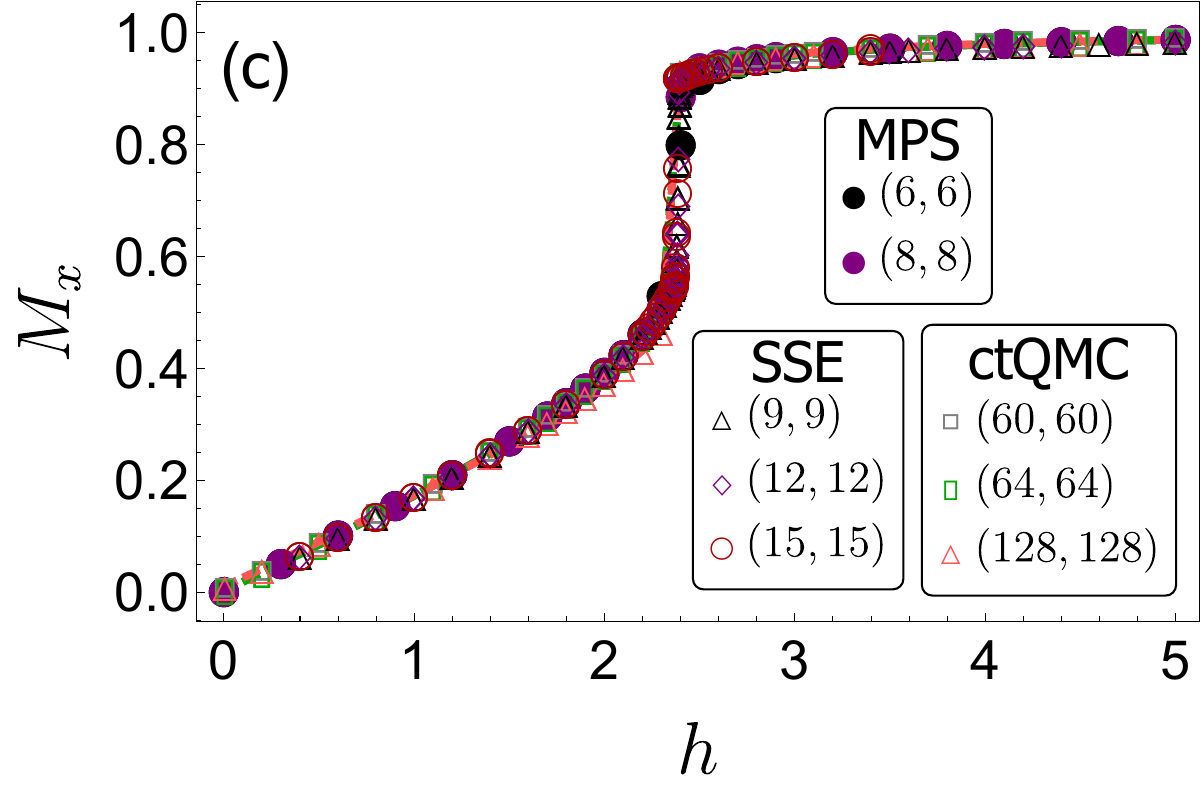}
   \end{subfigure}
   \begin{subfigure}[b]{0.3\textwidth}
      \includegraphics[width=\textwidth]{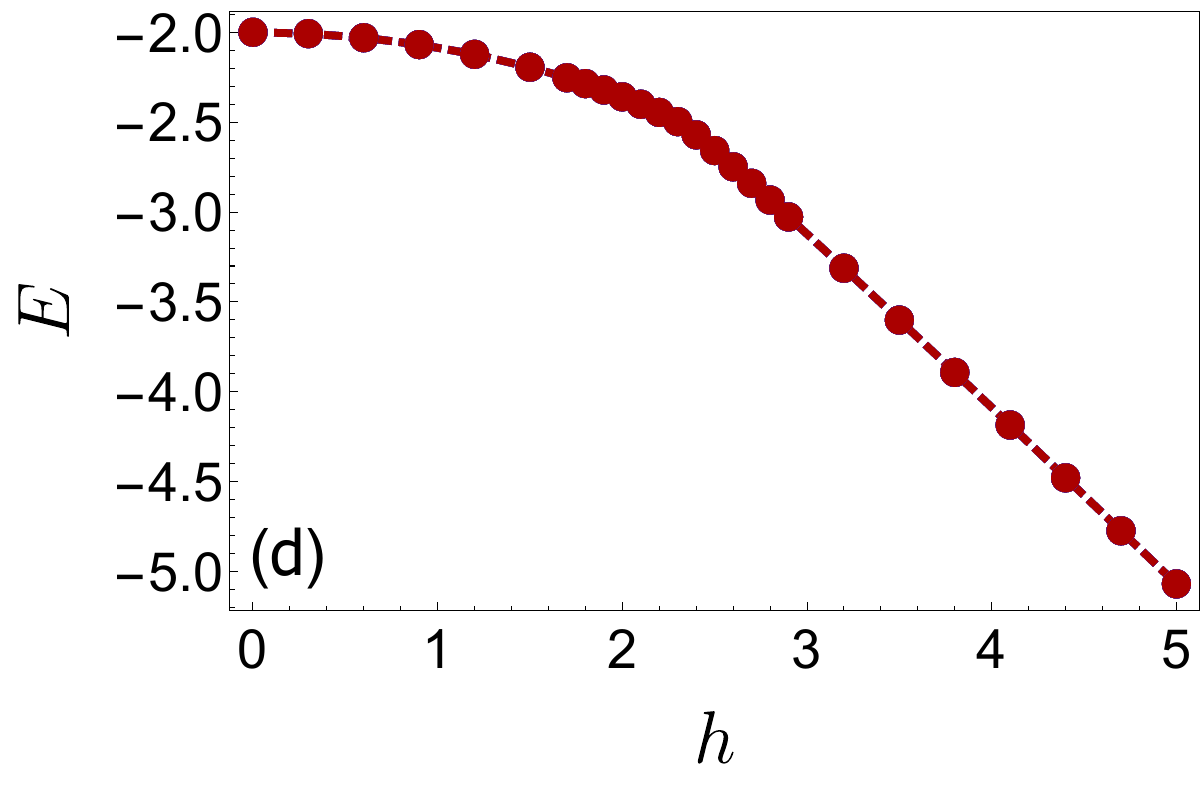}
   \end{subfigure}
   \begin{subfigure}[b]{0.3\textwidth}
     \includegraphics[width=\textwidth]{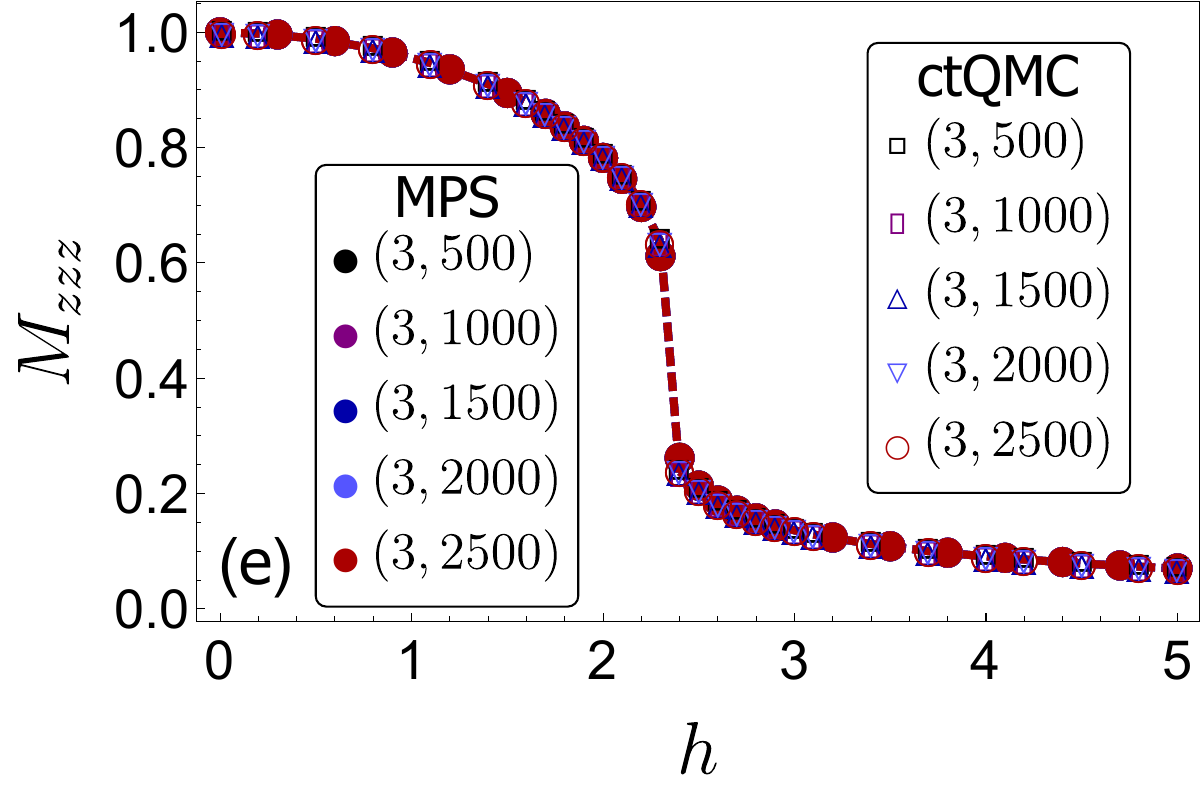}
   \end{subfigure}
   \begin{subfigure}[b]{0.3\textwidth}
   \includegraphics[width=\textwidth]{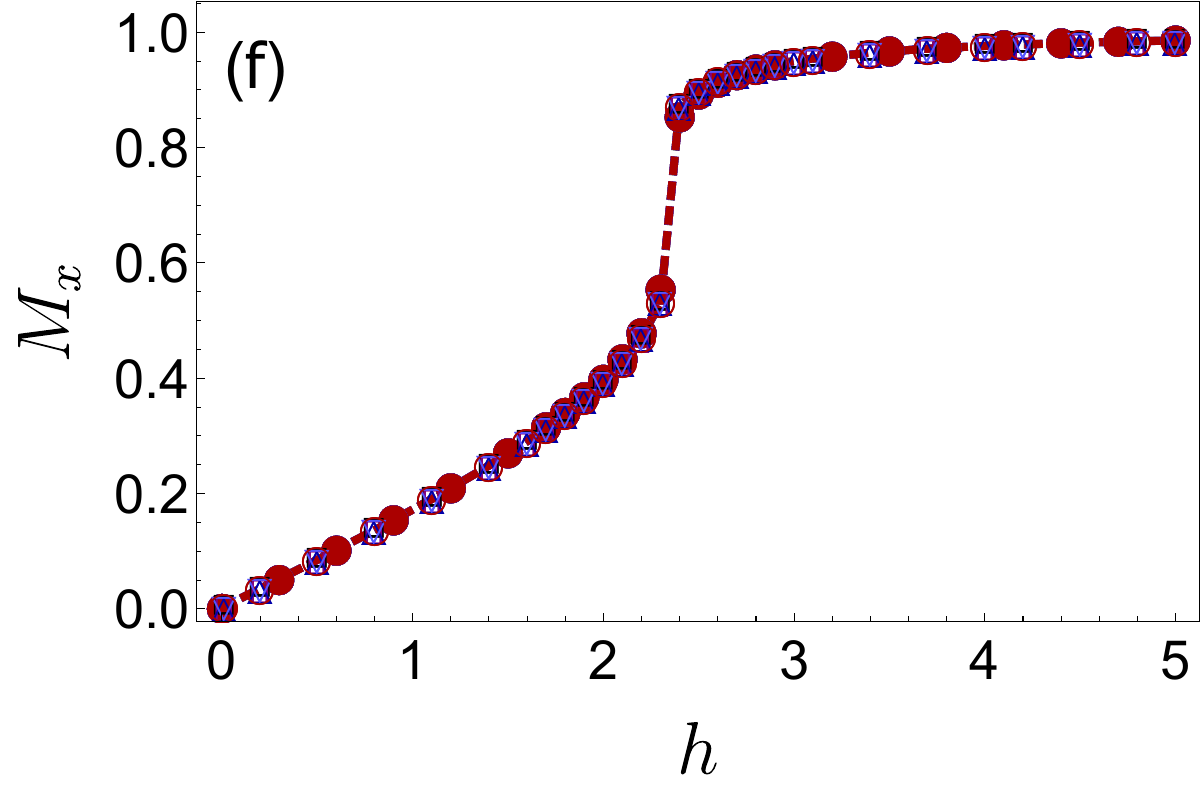}
   \end{subfigure}
   \caption{
      {\bf Phase transition of $\opcatBW{H}$ for PBC.}
      (a-c) Normalized ground state energy, average three-spin correlator for the interaction from Rule 60 and transverse magnetization as a function of $h$ for PBC and square systems sizes, as shown in panel (b). We compare our MPS and ctQMC results to those of Ref.~\cite{capponi2014baxter-wu} (indicated as SSE) in panel (c) for system sizes as shown in this panel.
      (d-f) Same but for quasi-1D geometries, as indicated in panel (e).
      }
    \label{fig:qBW}
\end{figure*}

In Fig.~\ref{fig:qBW} we show our numerics for the quantum Baxter-Wu model. Our results from numerical MPS and ctQMC simulations also suggest a quantum phase transition for the same value of the transverse field. In panels (a-c) we show square geometries, where our results coincide with those of Ref.~\cite{capponi2014baxter-wu}. 
Panels (d-f) show the same for strip geometries, for which the transition, while weaker, still seems first-order. In the context of our previous analysis, we expect that there will be additional SSB of the classical symmetries for the system sizes that possess multiple classical ground states according to the two counter-propagating CA prescription of Sec.~\ref{BWclassical}.

\section{Conclusions}{\label{conclusions}}

In this work we have presented a classification of 
$(d+1)$-dimensional plaquette spin models from the properties of the trajectories of the associated $d$-dimensional CA. 
Given an elementary CA in $d$ dimensions we have defined the corresponding simplest classical spin model whose energy is minimized by the trajectories of the corresponding CA. For each elementary 1D CA we have provided the associated classical spin model. The set of these $256$ models is built from eight fundamental models, each defined by a single kind of interaction, and we focused mostly on the model corresponding to Rule 150. For the case of 2D CA we have considered the rule that gives rise to the SPyM, a 3D generalization of the TPM. 

Endowing these plaquette models with a transverse field, we provided evidence for the existence and for the nature of their ground state quantum phase transitions based on a range of numerical techniques.
We then studied the BW model with the use of two counterpropagating CA. This technique allowed us to uncover the hidden SSB which accompanies its first-order quantum phase transition.

Beyond the specific models we considered here, our approach is general and applicable to CA of any neighborhood, in any finite field and in any dimension. Generalizations of the spin models we considered that are amenable to a similar treatment include those of Ref.~\cite{biswas2022beyond}, but also of the rest of the models of Ref.~\cite{heringa1989phase}. 

The CA we studied here are all defined in terms of synchronous evolution. 
Natural generalizations would be to consider CA with dynamics analogous to that of 
brickwork arrangements as in quantum circuit models 
\cite{bobenko1993on-two-integrable,inoue2018two-extensions,bertini2019entanglement,gopalakrishnan2019unitary,wilkinson2020exact,buca2021rule,
gombor2022superintegrable,wilkinson2022exact-solution,klobas2023exact} or probabilistic CA \cite{1985_Grinstein,1990_Lebowitz,2022_Hartarsky}.
Also, we note that some of the models described in Appx.~\ref{appendix:all-models} were also found recently in the context of measurement-induced entanglement phase transitions for random unitary circuits with dissipation \cite{li2023statistical}. 
Lastly, the study of these models at finite temperature consists of another interesting direction: 
the classical models might display glassy dynamics with emergent kinetic constraints, while the finite temperature phase diagram of the respective quantum models might be nontrivial and reveal interesting new phenomena.

\begin{acknowledgments}
  We are grateful to S. Balasubramanian, J.C\^{o}t\'{e}, A. Fahimniya, J. Lahtonen, E.Lake, C. Li, A. Gammon-Smith, N. Tandivasadakarn, M. Tikhanovskaya and M. Will for various, helpful discussions and S. Capponi for the data of Ref.~\cite{capponi2014baxter-wu} in Fig.~\ref{fig:qBW}.
  We acknowledge financial support from EPSRC Grant no.\ EP/R04421X/1, the Leverhulme Trust Grant No. RPG-2018-181, and University of Nottingham grant no.\ FiF1/3.
  LC was supported by an EPSRC Doctoral prize from the University of Nottingham.
  Simulations were performed using the University of Nottingham Augusta HPC cluster, and the Sulis Tier 2 HPC platform hosted by the Scientific Computing Research Technology Platform at the University of Warwick (funded by EPSRC Grant EP/T022108/1 and the HPC Midlands+ consortium). 
  \end{acknowledgments}

\balance

\appendix
\section{1D elementary CA and their 2D parent Hamiltonians}{\label{appendix:all-models}}

Here we present the simplest classical ``parent'' Hamiltonians for the 256 elementary CA \cite{auerbach_interacting_1994} for which the cycles of the CA are minimum energy configurations. The relation between a CA rule and the corresponding classical Hamiltonian is given in Tables~\ref{tab-0}-\ref{tab-240}. The tables show the CA rule number, the update rule, the local interactions that define the spin model energy function, and the coefficients of each interaction term of Eqs.~\eqref{m0}-\eqref{m150} that form the linear combination of the given models. 
Specifically, the energy function for CA rule $\mu$ is a sum of local interaction terms
\begin{equation}
    E_\mu = \sum_{\{p,q,r,s\}}^N \varepsilon_{p,q,r,s}^{(\mu)}, 
\end{equation}
where $\{p,q,r,s\}$ indicate the spins in the neighbourhood like that of Fig.~\ref{fig:r150}. Our choice of spin models is such that, for linear CA, there is one interaction term per unit cell so that there is an equal total number of interaction terms and spins on the lattice, $N$. In general, the interaction $\varepsilon_{p,q,r,s}^{(\mu)}$ can be written in terms of the interactions of the irreducible models \eqref{m0}-\eqref{m150}:
\begin{equation}
    \varepsilon_{p,q,r,s}^{(\mu)} = 
        \sum_{\nu \in {\rm \, Irr}}
        \alpha_\nu^{(\mu)} \varepsilon_{p,q,r,s}^{(\nu)}
\end{equation}
where ``Irr'' are the eight irreducible models of Eqs.~\eqref{m0}-\eqref{m150}. 

In the table we also use the expressions:
\begin{align}
    &\text{CZ}_{a, b} = \frac{1}{2}(1 + \sigma_a + \sigma_b - \sigma_a \sigma_b) \\
    &\text{CCZ}_{a, b, c} = 1 - \frac{1}{4} (1 - \sigma_a - \sigma_b - \sigma_c  
    \nonumber \\
    & \;\;\;\;\;\;\;\;\;\;
    + \sigma_a \sigma_b + \sigma_a \sigma_c + \sigma_b \sigma_c
    - \sigma_a \sigma_b \sigma_c ).
\end{align}

\begin{widetext}

\begin{table}
    \[
\begin{array}{ccccccccccc}
            \hline
            \text{CA Rule no.} & \text{Update rule} & \varepsilon_{p,q,r,s}^{(\mu)}  & \alpha_0 & \alpha_{240} & \alpha_{204} & \alpha_{170} & \alpha_{60} & \alpha_{90} & \alpha_{102} & \alpha_{150} \\
            \hline
            0 & 0 & -\sigma_s & 1 & 0 & 0 & 0 & 0 & 0 & 0 & 0 \\
            1 & 1+p+q+p q+r+p r+q r+p q r & \sigma_p \sigma_q \sigma_r \sigma_s \text{CZ}_{p,q} \text{CZ}_{p,r} \text{CZ}_{q,r} \text{CCZ}_{p,q,r} & \frac{3}{4} & -\frac{1}{4} & -\frac{1}{4} & -\frac{1}{4} & -\frac{1}{4} & -\frac{1}{4} & -\frac{1}{4} & -\frac{1}{4} \\
            2 & r+p r+q r+p q r & -\sigma_r \sigma_s \text{CZ}_{p,r} \text{CZ}_{q,r} \text{CCZ}_{p,q,r} & \frac{3}{4} & -\frac{1}{4} & -\frac{1}{4} & \frac{1}{4} & -\frac{1}{4} & \frac{1}{4} & \frac{1}{4} & \frac{1}{4} \\
            3 & 1+p+q+p q & \sigma_p \sigma_q \sigma_s \text{CZ}_{p,q} & \frac{1}{2} & -\frac{1}{2} & -\frac{1}{2} & 0 & -\frac{1}{2} & 0 & 0 & 0 \\
            4 & q+p q+q r+p q r & -\sigma_q \sigma_s \text{CZ}_{p,q} \text{CZ}_{q,r} \text{CCZ}_{p,q,r} & \frac{3}{4} & -\frac{1}{4} & \frac{1}{4} & -\frac{1}{4} & \frac{1}{4} & -\frac{1}{4} & \frac{1}{4} & \frac{1}{4} \\
            5 & 1+p+r+p r & \sigma_p \sigma_r \sigma_s \text{CZ}_{p,r} & \frac{1}{2} & -\frac{1}{2} & 0 & -\frac{1}{2} & 0 & -\frac{1}{2} & 0 & 0 \\
            6 & q+p q+r+p r & -\sigma_q \sigma_r \sigma_s \text{CZ}_{p,q} \text{CZ}_{p,r} & \frac{1}{2} & -\frac{1}{2} & 0 & 0 & 0 & 0 & \frac{1}{2} & \frac{1}{2} \\
\hline
        \end{array}
    \]
    \caption{2D classical spin models from 1D CA Rules 0 to 39.}
    \label{tab-0}
\end{table}

\begin{table}
    \[
        \begin{array}{ccccccccccc}
                \hline
                \text{CA Rule no.} & \text{Update rule} & \varepsilon_{p,q,r,s}^{(\mu)}  & \alpha_0 & \alpha_{240} & \alpha_{204} & \alpha_{170} & \alpha_{60} & \alpha_{90} & \alpha_{102} & \alpha_{150} \\
                \hline
                40 & p r+q r & -\sigma_s \text{CZ}_{p,r} \text{CZ}_{q,r} & \frac{1}{2} & 0 & 0 & \frac{1}{2} & \frac{1}{2} & 0 & 0 & -\frac{1}{2} \\
                41 & 1+p+q+p q+r+p q r & \sigma_p \sigma_q \sigma_r \sigma_s \text{CZ}_{p,q} \text{CCZ}_{p,q,r} & \frac{1}{4} & -\frac{1}{4} & -\frac{1}{4} & \frac{1}{4} & \frac{1}{4} & -\frac{1}{4} & -\frac{1}{4} & -\frac{3}{4} \\
                42 & r+p q r & -\sigma_r \sigma_s \text{CCZ}_{p,q,r} & \frac{1}{4} & -\frac{1}{4} & -\frac{1}{4} & \frac{3}{4} & \frac{1}{4} & \frac{1}{4} & \frac{1}{4} & -\frac{1}{4} \\
                43 & 1+p+q+p q+p r+q r & \sigma_p \sigma_q \sigma_s \text{CZ}_{p,q} \text{CZ}_{p,r} \text{CZ}_{q,r} & 0 & -\frac{1}{2} & -\frac{1}{2} & \frac{1}{2} & 0 & 0 & 0 & -\frac{1}{2} \\
                44 & q+p q+p r+p q r & -\sigma_q \sigma_s \text{CZ}_{p,q} \text{CZ}_{p,r} \text{CCZ}_{p,q,r} & \frac{1}{4} & -\frac{1}{4} & \frac{1}{4} & \frac{1}{4} & \frac{3}{4} & -\frac{1}{4} & \frac{1}{4} & -\frac{1}{4} \\
                45 & 1+p+r+q r & \sigma_p \sigma_r \sigma_s \text{CZ}_{q,r} & 0 & -\frac{1}{2} & 0 & 0 & \frac{1}{2} & -\frac{1}{2} & 0 & -\frac{1}{2} \\
                46 & q+p q+r+q r & -\sigma_q \sigma_r \sigma_s \text{CZ}_{p,q} \text{CZ}_{q,r} & 0 & -\frac{1}{2} & 0 & \frac{1}{2} & \frac{1}{2} & 0 & \frac{1}{2} & 0 \\
\hline
        \end{array}      
    \]
    \caption{2D classical spin models from 1D CA Rules 40 to 79.}
    \label{tab-40}
\end{table}

\begin{table}
    \[
        \begin{array}{ccccccccccc}
            \hline
            \text{CA Rule no.} & \text{Update rule} & \varepsilon_{p,q,r,s}^{(\mu)}  & \alpha_0 & \alpha_{240} & \alpha_{204} & \alpha_{170} & \alpha_{60} & \alpha_{90} & \alpha_{102} & \alpha_{150} \\
            \hline
            80 & p+p r & -\sigma_p \sigma_s \text{CZ}_{p,r} & \frac{1}{2} & \frac{1}{2} & 0 & -\frac{1}{2} & 0 & \frac{1}{2} & 0 & 0 \\
            81 & 1+q+p q+r+q r+p q r & \sigma_q \sigma_r \sigma_s \text{CZ}_{p,q} \text{CZ}_{q,r} \text{CCZ}_{p,q,r} & \frac{1}{4} & \frac{1}{4} & -\frac{1}{4} & -\frac{3}{4} & -\frac{1}{4} & \frac{1}{4} & -\frac{1}{4} & -\frac{1}{4} \\
            82 & p+r+q r+p q r & -\sigma_p \sigma_r \sigma_s \text{CZ}_{q,r} \text{CCZ}_{p,q,r} & \frac{1}{4} & \frac{1}{4} & -\frac{1}{4} & -\frac{1}{4} & -\frac{1}{4} & \frac{3}{4} & \frac{1}{4} & \frac{1}{4} \\
            83 & 1+q+p q+p r & \sigma_q \sigma_s \text{CZ}_{p,q} \text{CZ}_{p,r} & 0 & 0 & -\frac{1}{2} & -\frac{1}{2} & -\frac{1}{2} & \frac{1}{2} & 0 & 0 \\
            84 & p+q+p q+p r+q r+p q r & -\sigma_p \sigma_q \sigma_s \text{CZ}_{p,q} \text{CZ}_{p,r} \text{CZ}_{q,r} \text{CCZ}_{p,q,r} & \frac{1}{4} & \frac{1}{4} & \frac{1}{4} & -\frac{3}{4} & \frac{1}{4} & \frac{1}{4} & \frac{1}{4} & \frac{1}{4} \\
            85 & 1+r & \sigma_r \sigma_s & 0 & 0 & 0 & -1 & 0 & 0 & 0 & 0 \\
            86 & p+q+p q+r & -\sigma_p \sigma_q \sigma_r \sigma_s \text{CZ}_{p,q} & 0 & 0 & 0 & -\frac{1}{2} & 0 & \frac{1}{2} & \frac{1}{2} & \frac{1}{2} \\
\hline
        \end{array}
    \]
    \caption{2D classical spin models from 1D CA Rules 80 to 119.}
    \label{tab-80}
\end{table}

\begin{table}
    \[
        \begin{array}{ccccccccccc}
                \hline
                \text{CA Rule no.} & \text{Update rule} & \varepsilon_{p,q,r,s}^{(\mu)}  & \alpha_0 & \alpha_{240} & \alpha_{204} & \alpha_{170} & \alpha_{60} & \alpha_{90} & \alpha_{102} & \alpha_{150} \\
                \hline
                120 & p+q r & -\sigma_p \sigma_s \text{CZ}_{q,r} & 0 & \frac{1}{2} & 0 & 0 & \frac{1}{2} & \frac{1}{2} & 0 & -\frac{1}{2} \\
                121 & 1+q+p q+r+p r+p q r & \sigma_q \sigma_r \sigma_s \text{CZ}_{p,q} \text{CZ}_{p,r} \text{CCZ}_{p,q,r} & -\frac{1}{4} & \frac{1}{4} & -\frac{1}{4} & -\frac{1}{4} & \frac{1}{4} & \frac{1}{4} & -\frac{1}{4} & -\frac{3}{4} \\
                122 & p+r+p r+p q r & -\sigma_p \sigma_r \sigma_s \text{CZ}_{p,r} \text{CCZ}_{p,q,r} & -\frac{1}{4} & \frac{1}{4} & -\frac{1}{4} & \frac{1}{4} & \frac{1}{4} & \frac{3}{4} & \frac{1}{4} & -\frac{1}{4} \\
                123 & 1+q+p q+q r & \sigma_q \sigma_s \text{CZ}_{p,q} \text{CZ}_{q,r} & -\frac{1}{2} & 0 & -\frac{1}{2} & 0 & 0 & \frac{1}{2} & 0 & -\frac{1}{2} \\
                124 & p+q+p q+p q r & -\sigma_p \sigma_q \sigma_s \text{CZ}_{p,q} \text{CCZ}_{p,q,r} & -\frac{1}{4} & \frac{1}{4} & \frac{1}{4} & -\frac{1}{4} & \frac{3}{4} & \frac{1}{4} & \frac{1}{4} & -\frac{1}{4} \\
                125 & 1+r+p r+q r & \sigma_r \sigma_s \text{CZ}_{p,r} \text{CZ}_{q,r} & -\frac{1}{2} & 0 & 0 & -\frac{1}{2} & \frac{1}{2} & 0 & 0 & -\frac{1}{2} \\
                126 & p+q+p q+r+p r+q r & -\sigma_p \sigma_q \sigma_r \sigma_s \text{CZ}_{p,q} \text{CZ}_{p,r} \text{CZ}_{q,r} & -\frac{1}{2} & 0 & 0 & 0 & \frac{1}{2} & \frac{1}{2} & \frac{1}{2} & 0 \\
\hline
        \end{array}
    \]
    \caption{2D classical spin models from 1D CA Rules 120 to 159.}
    \label{tab-120}
\end{table}

\begin{table}
    \[
        \begin{array}{ccccccccccc}
            \hline
            \text{CA Rule no.} & \text{Update rule} & \varepsilon_{p,q,r,s}^{(\mu)}  & \alpha_0 & \alpha_{240} & \alpha_{204} & \alpha_{170} & \alpha_{60} & \alpha_{90} & \alpha_{102} & \alpha_{150} \\
            \hline
            160 & p r & -\sigma_s \text{CZ}_{p,r} & \frac{1}{2} & \frac{1}{2} & 0 & \frac{1}{2} & 0 & -\frac{1}{2} & 0 & 0 \\
            161 & 1+p+q+p q+r+q r+p q r & \sigma_p \sigma_q \sigma_r \sigma_s \text{CZ}_{p,q} \text{CZ}_{q,r} \text{CCZ}_{p,q,r} & \frac{1}{4} & \frac{1}{4} & -\frac{1}{4} & \frac{1}{4} & -\frac{1}{4} & -\frac{3}{4} & -\frac{1}{4} & -\frac{1}{4} \\
            162 & r+q r+p q r & -\sigma_r \sigma_s \text{CZ}_{q,r} \text{CCZ}_{p,q,r} & \frac{1}{4} & \frac{1}{4} & -\frac{1}{4} & \frac{3}{4} & -\frac{1}{4} & -\frac{1}{4} & \frac{1}{4} & \frac{1}{4} \\
            163 & 1+p+q+p q+p r & \sigma_p \sigma_q \sigma_s \text{CZ}_{p,q} \text{CZ}_{p,r} & 0 & 0 & -\frac{1}{2} & \frac{1}{2} & -\frac{1}{2} & -\frac{1}{2} & 0 & 0 \\
            164 & q+p q+p r+q r+p q r & -\sigma_q \sigma_s \text{CZ}_{p,q} \text{CZ}_{p,r} \text{CZ}_{q,r} \text{CCZ}_{p,q,r} & \frac{1}{4} & \frac{1}{4} & \frac{1}{4} & \frac{1}{4} & \frac{1}{4} & -\frac{3}{4} & \frac{1}{4} & \frac{1}{4} \\
            165 & 1+p+r & \sigma_p \sigma_r \sigma_s & 0 & 0 & 0 & 0 & 0 & -1 & 0 & 0 \\
            166 & q+p q+r & -\sigma_q \sigma_r \sigma_s \text{CZ}_{p,q} & 0 & 0 & 0 & \frac{1}{2} & 0 & -\frac{1}{2} & \frac{1}{2} & \frac{1}{2} \\
\hline
        \end{array}
    \]
    \caption{2D classical spin models from 1D CA Rules 160 to 199.}
    \label{tab-160}
\end{table}

\begin{table}
    \[
        \begin{array}{ccccccccccc}
            \hline
            \text{CA Rule no.} & \text{Update rule} & \varepsilon_{p,q,r,s}^{(\mu)}  & \alpha_0 & \alpha_{240} & \alpha_{204} & \alpha_{170} & \alpha_{60} & \alpha_{90} & \alpha_{102} & \alpha_{150} \\
            \hline
            200 & p q+q r+p q r & -\sigma_s \text{CZ}_{p,q} \text{CZ}_{q,r} \text{CCZ}_{p,q,r} & \frac{1}{4} & \frac{1}{4} & \frac{3}{4} & \frac{1}{4} & -\frac{1}{4} & \frac{1}{4} & -\frac{1}{4} & -\frac{1}{4} \\
            201 & 1+p+q+r+p r & \sigma_p \sigma_q \sigma_r \sigma_s \text{CZ}_{p,r} & 0 & 0 & \frac{1}{2} & 0 & -\frac{1}{2} & 0 & -\frac{1}{2} & -\frac{1}{2} \\
            202 & p q+r+p r & -\sigma_r \sigma_s \text{CZ}_{p,q} \text{CZ}_{p,r} & 0 & 0 & \frac{1}{2} & \frac{1}{2} & -\frac{1}{2} & \frac{1}{2} & 0 & 0 \\
            203 & 1+p+q+q r+p q r & \sigma_p \sigma_q \sigma_s \text{CZ}_{q,r} \text{CCZ}_{p,q,r} & -\frac{1}{4} & -\frac{1}{4} & \frac{1}{4} & \frac{1}{4} & -\frac{3}{4} & \frac{1}{4} & -\frac{1}{4} & -\frac{1}{4} \\
            204 & q & -\sigma_q \sigma_s & 0 & 0 & 1 & 0 & 0 & 0 & 0 & 0 \\
            205 & 1+p+p q+r+p r+q r+p q r & \sigma_p \sigma_r \sigma_s \text{CZ}_{p,q} \text{CZ}_{p,r} \text{CZ}_{q,r} \text{CCZ}_{p,q,r} & -\frac{1}{4} & -\frac{1}{4} & \frac{3}{4} & -\frac{1}{4} & -\frac{1}{4} & -\frac{1}{4} & -\frac{1}{4} & -\frac{1}{4} \\
            206 & q+r+p r+q r+p q r & -\sigma_q \sigma_r \sigma_s \text{CZ}_{p,r} \text{CZ}_{q,r} \text{CCZ}_{p,q,r} & -\frac{1}{4} & -\frac{1}{4} & \frac{3}{4} & \frac{1}{4} & -\frac{1}{4} & \frac{1}{4} & \frac{1}{4} & \frac{1}{4} \\
\hline
        \end{array}
    \]
    \caption{2D classical spin models from 1D CA Rules 200 to 239.}
    \label{tab-200}
\end{table}

\begin{table}
    \[
        \begin{array}{ccccccccccc}
            \hline
            \text{CA Rule no.} & \text{Update rule} & \varepsilon_{p,q,r,s}^{(\mu)}  & \alpha_0 & \alpha_{240} & \alpha_{204} & \alpha_{170} & \alpha_{60} & \alpha_{90} & \alpha_{102} & \alpha_{150} \\
            \hline
            240 & p & -\sigma_p \sigma_s & 0 & 1 & 0 & 0 & 0 & 0 & 0 & 0 \\
            241 & 1+q+p q+r+p r+q r+p q r & \sigma_q \sigma_r \sigma_s \text{CZ}_{p,q} \text{CZ}_{p,r} \text{CZ}_{q,r} \text{CCZ}_{p,q,r} & -\frac{1}{4} & \frac{3}{4} & -\frac{1}{4} & -\frac{1}{4} & -\frac{1}{4} & -\frac{1}{4} & -\frac{1}{4} & -\frac{1}{4} \\
            242 & p+r+p r+q r+p q r & -\sigma_p \sigma_r \sigma_s \text{CZ}_{p,r} \text{CZ}_{q,r} \text{CCZ}_{p,q,r} & -\frac{1}{4} & \frac{3}{4} & -\frac{1}{4} & \frac{1}{4} & -\frac{1}{4} & \frac{1}{4} & \frac{1}{4} & \frac{1}{4} \\
            243 & 1+q+p q & \sigma_q \sigma_s \text{CZ}_{p,q} & -\frac{1}{2} & \frac{1}{2} & -\frac{1}{2} & 0 & -\frac{1}{2} & 0 & 0 & 0 \\
            244 & p+q+p q+q r+p q r & -\sigma_p \sigma_q \sigma_s \text{CZ}_{p,q} \text{CZ}_{q,r} \text{CCZ}_{p,q,r} & -\frac{1}{4} & \frac{3}{4} & \frac{1}{4} & -\frac{1}{4} & \frac{1}{4} & -\frac{1}{4} & \frac{1}{4} & \frac{1}{4} \\
            245 & 1+r+p r & \sigma_r \sigma_s \text{CZ}_{p,r} & -\frac{1}{2} & \frac{1}{2} & 0 & -\frac{1}{2} & 0 & -\frac{1}{2} & 0 & 0 \\
            246 & p+q+p q+r+p r & -\sigma_p \sigma_q \sigma_r \sigma_s \text{CZ}_{p,q} \text{CZ}_{p,r} & -\frac{1}{2} & \frac{1}{2} & 0 & 0 & 0 & 0 & \frac{1}{2} & \frac{1}{2} \\
\hline
        \end{array}
    \]
    \caption{2D classical spin models from 1D CA Rules 240 to 255.}
    \label{tab-240}
\end{table}

\end{widetext}
\section{Other CA and their spin models}{\label{appendix:other-CA}}

Here we discuss the extension of the class of spin models related to CA beyond those presented in the main text. Allowing for a larger neighbourhood that defines the given CA rule provides significant freedom in defining models with specific properties, e.g. models with specific system size sequences that exhibit an exact spin to defect duality, as in Ref.~\cite{newman1999glassy}. Specifically, this last property gets reflected to a sequence of system sizes in the finite size scaling of the respective spin model where the classical ground state degeneracy of the classical model is always equal to one. This in turn leads to the lack of any nontrivial symmetries of the quantum models and a first order quantum phase transition without the accompanying SSB.

This feature holds for linear CA which possess an $A$-matrix, as in Sec.~\ref{1d_CA}, which is a sum of the diagonal matrix and the left or right shift map (see further details in Ref.~\cite{sfairopoulos2023boundary}). By generalizing beyond elementary CA to CA with rules defined on larger neighbourhoods than that of Fig.~\ref{fig:r150}(a), one can obtain other CA with the above property and their corresponding 2D spin models. 

Consider an 1D CA with a larger neighbourhood of $n$ contiguous sites, where we label the sites (and their states) $(p_1, p_2, \cdots, p_n)$ [an elementary CA corresponds therefore to $n=3$, and $(p_1, p_2, p_3) = (p,q,r)$ in Fig.~\ref{fig:r150}(a)]. For example, for $n=5$, three possible rules with this neighbourhood are  
\begin{equation}
    s = p_1 + p_3, \; 
    s = p_1 + p_5, \; 
    s = p_3 + p_5 \;\; \mod{2}.
\end{equation}
These are related to 2D spin models with a neighbourhood of $2r+1$ sites with $r=2$ and share the property of a unique minimum energy configuration for certain system sizes. 

Similar generalizations to Rule 60 in 3D can be obtained for Rules 90 and 150.
In terms of the classical spin models, for the 3D version of Eq.~\eqref{m90} we get
\begin{equation}
    E_{90}^{(3D)} 
    = 
    -\sum_{ \{ p_i\} }^N 
    \sigma_{p_1} 
    \sigma_{p_3} 
    \sigma_{p_7} 
    \sigma_{p_9} 
    \sigma_{s}, 
\end{equation}
where $\{p_1, \dots, p_9\}$ span the Moore neighbourhood and $N$ corresponds to the total number of lattice sites, $N = L \times M$ which is also equal to the interaection terms of the energy function. In turn, by generalizing Eq.~\eqref{m150} we get
\begin{equation}
    E_{150}^{(3D)} 
    =
    -\sum_{ \{ p_i \} }^N 
    \left(\prod_{i=1}^9 \sigma_{p_i} \right)\sigma_{s},
\end{equation}
where the product includes all spins in the Moore neighbourhood for the update of $\sigma_s$ and $N = K \times L \times M$ the number of lattice sites which is also equal to the number of interaction terms.
Similar generalizations can be constructed for CA that involve the update of site $s$ through the two previous row configurations, leading to matrix CA (see for example Ref.~\cite{biswas2022beyond}).
\section{Numerics for OBC}{\label{appendix:numerics}}

\begin{figure*}
    \centering
    \begin{subfigure}[b]{0.3\textwidth}
        \includegraphics[width=54mm, height=39mm]{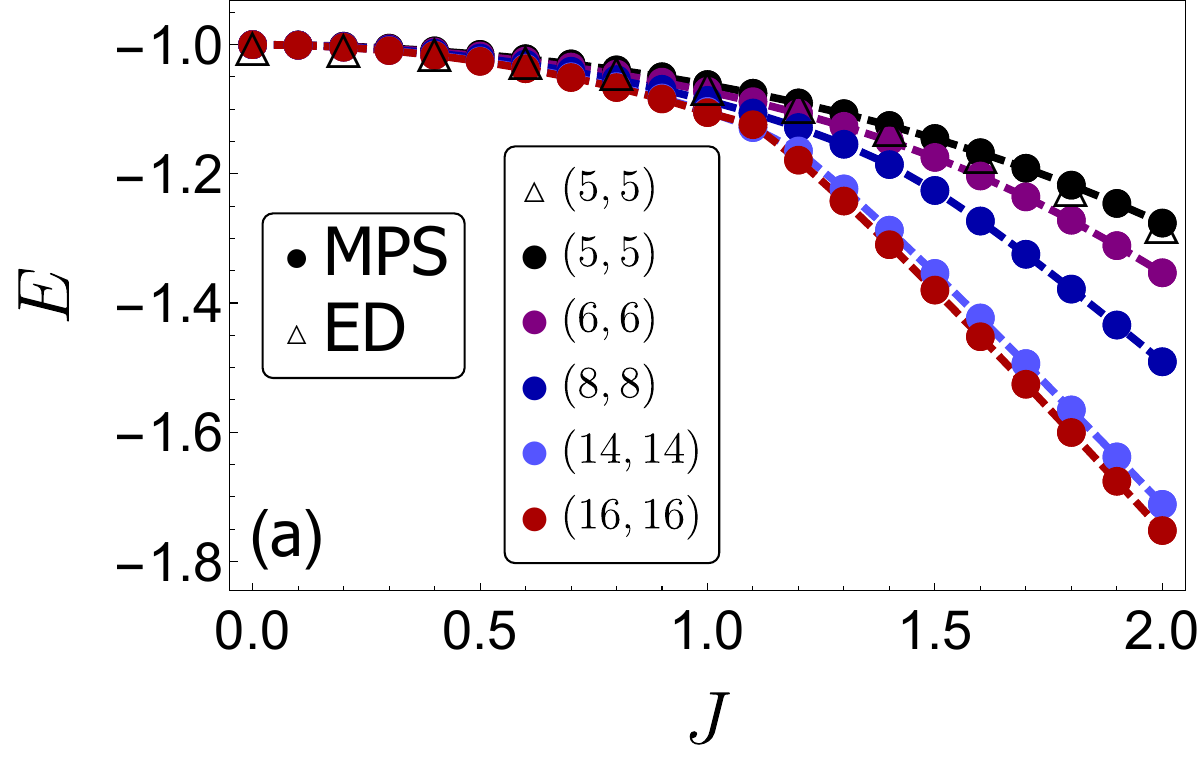}
     \end{subfigure}
     \begin{subfigure}[b]{0.3\textwidth}
        \includegraphics[width=54mm, height=39mm]{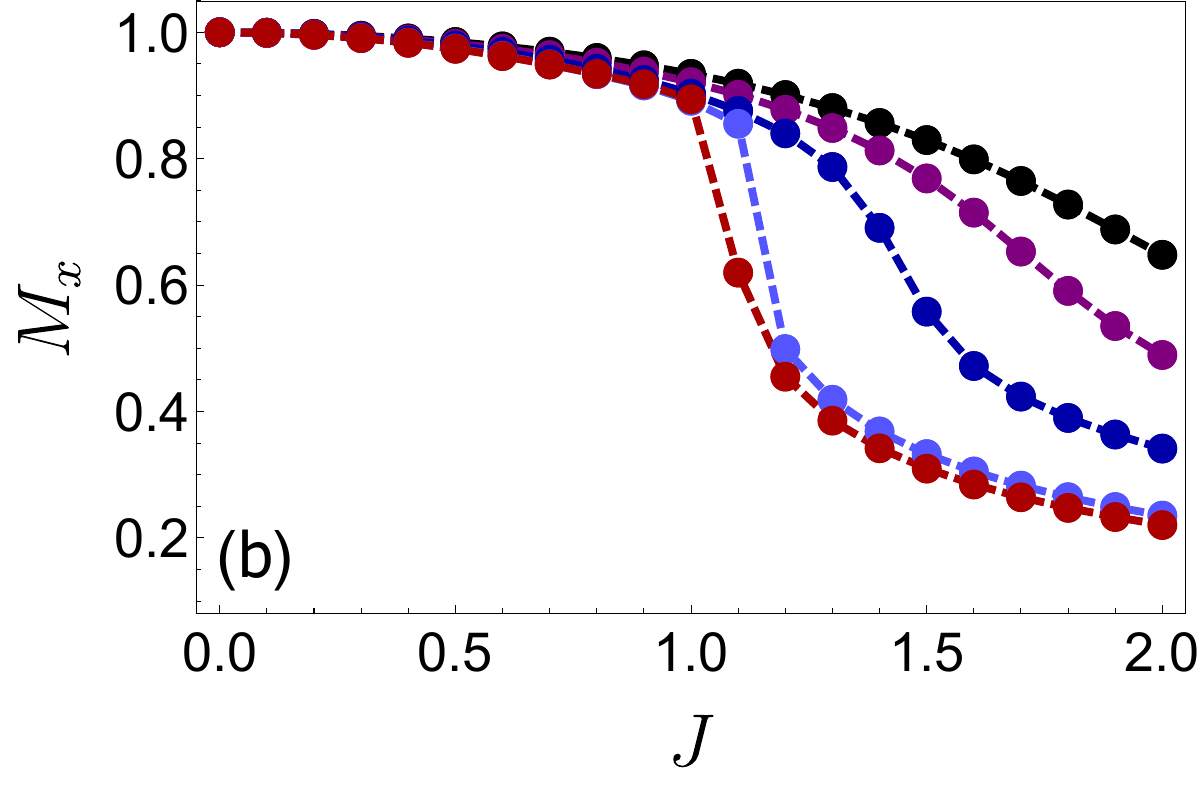}
     \end{subfigure}
     \begin{subfigure}[b]{0.3\textwidth}
       \includegraphics[width=54mm, height=39mm]{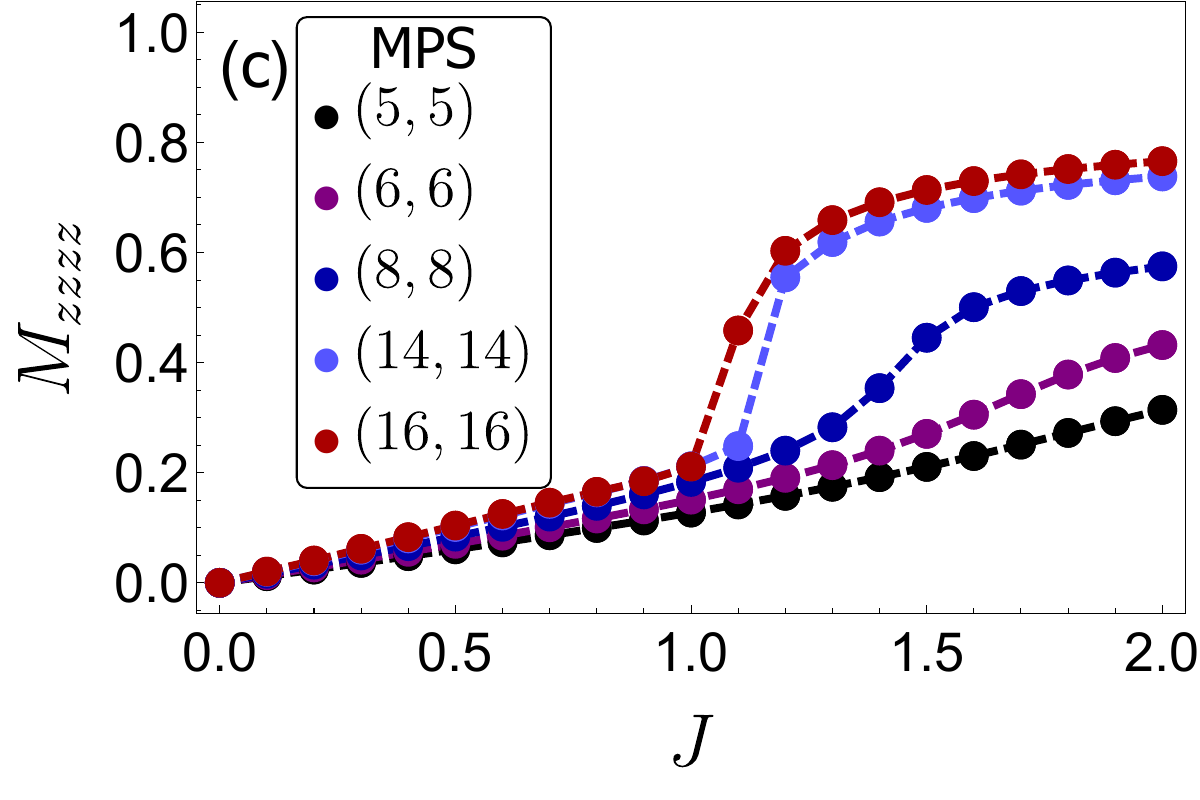}
    \end{subfigure}
    \begin{subfigure}[b]{0.3\textwidth}
      \includegraphics[width=54mm, height=39mm]{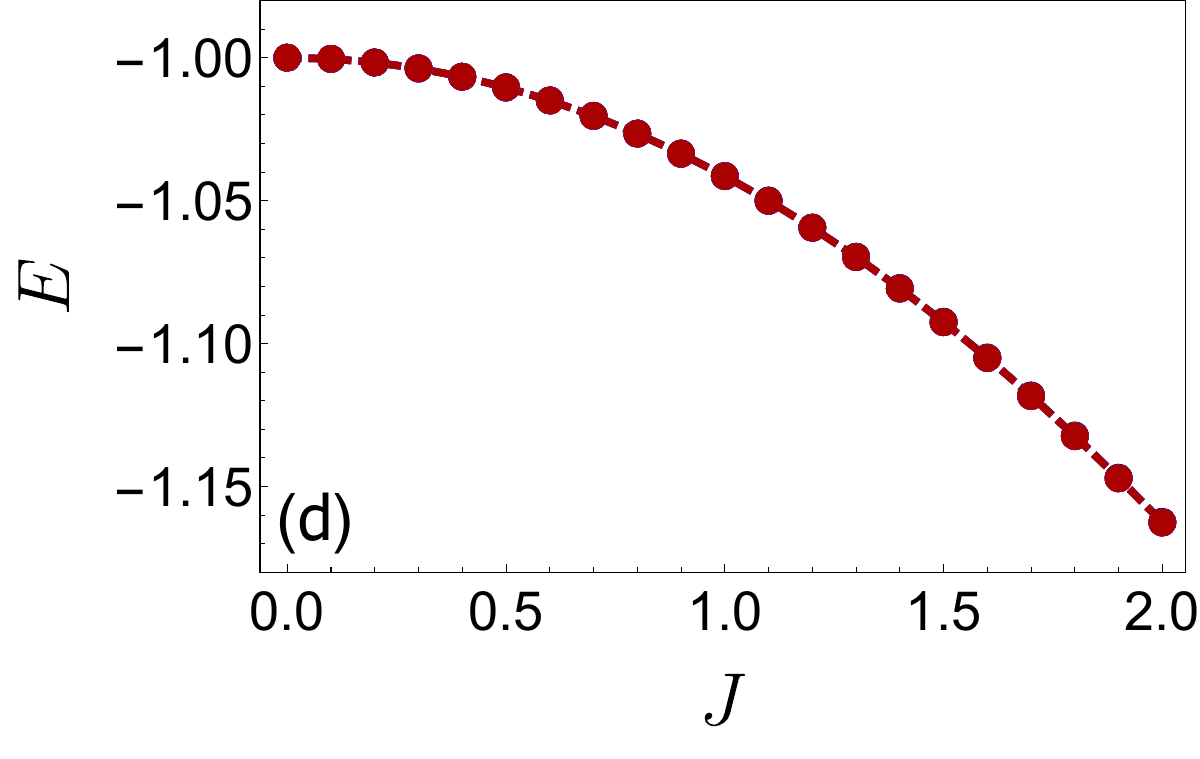}
   \end{subfigure}
   \begin{subfigure}[b]{0.3\textwidth}
      \includegraphics[width=54mm, height=39mm]{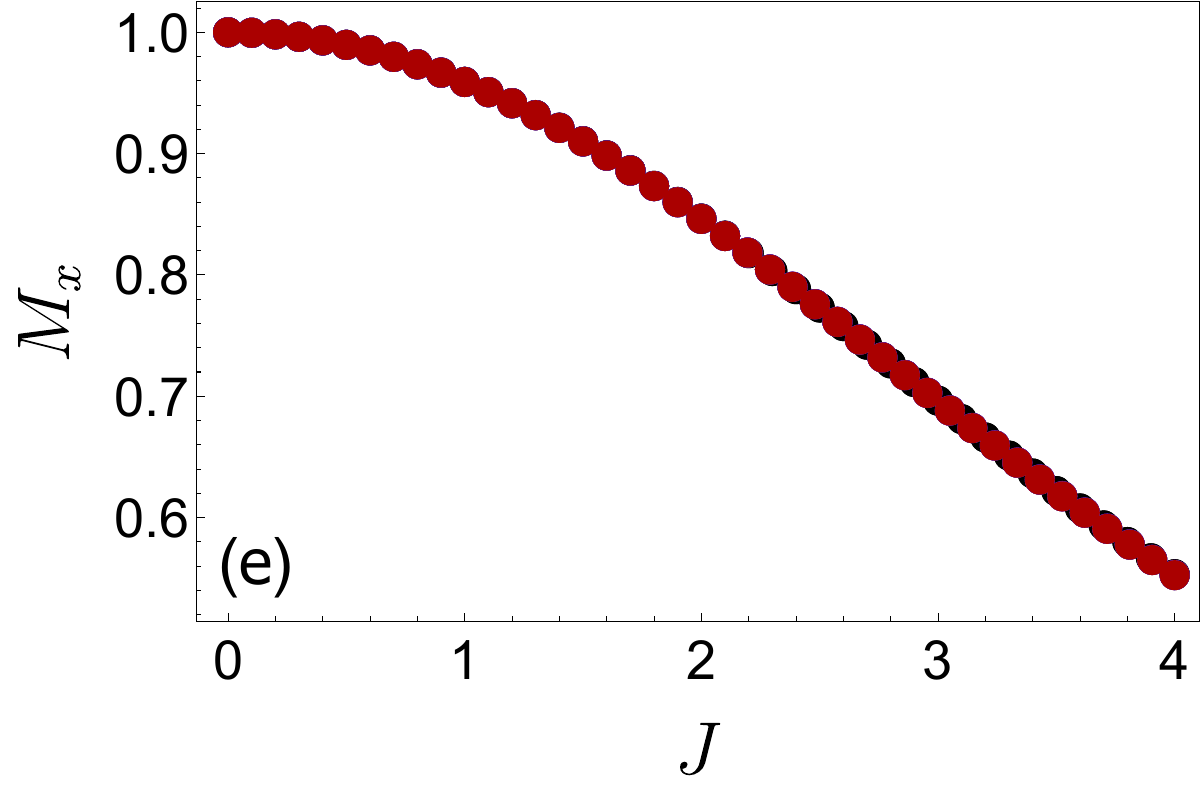}
   \end{subfigure}
   \begin{subfigure}[b]{0.3\textwidth}
     \includegraphics[width=54mm, height=39mm]{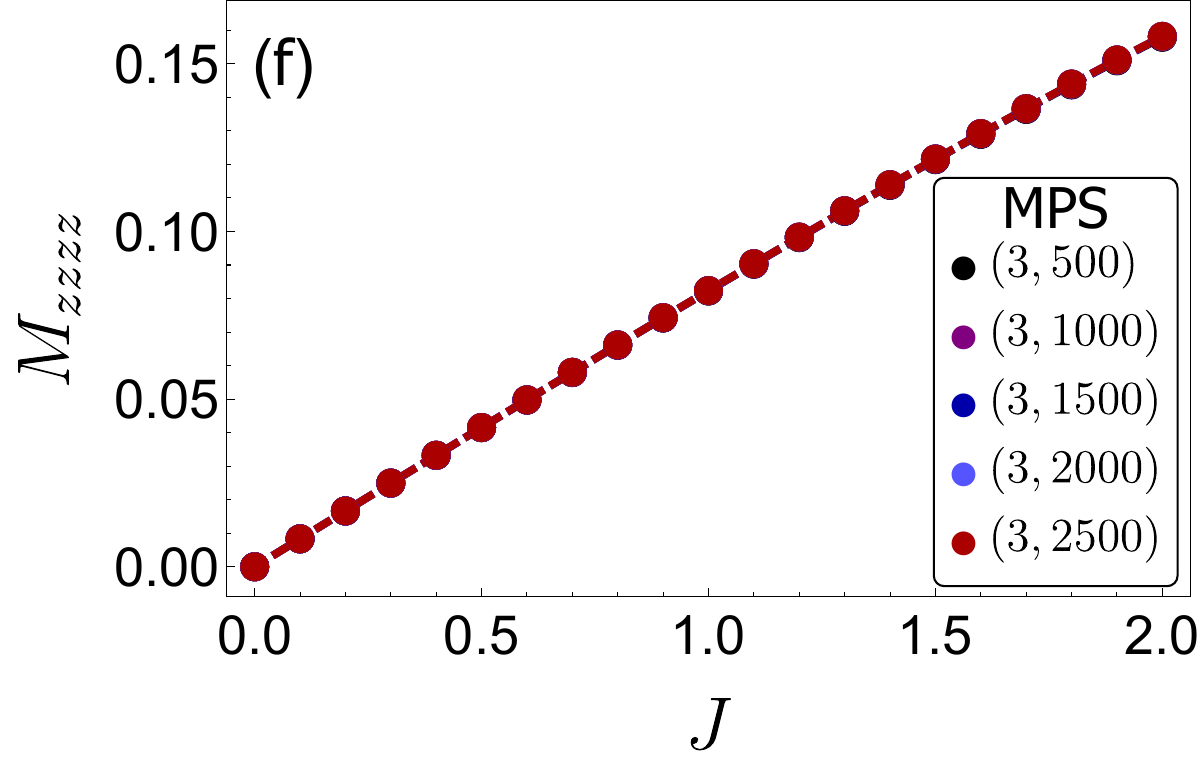}
  \end{subfigure}
     \caption[Rule 150, OBC, square]{
      {\bf Quantum phase transition of $H_{150}$ for OBC.}
     (a) The normalized by the system size ground state energy as a function of $J$. Empty symbols are from ED while filled symbols are from numerical MPS.
     (b) Transverse magnetization as a function of $J$. 
     (c) Average four-spin interaction as a function of $J$.
     (d-f) Similar to (a-c) but for the rectangular system sizes given in the inset of panel (f).}
     \label{fig:Rule_150_OBC}
\end{figure*}

\begin{figure*}
   \centering
   \begin{subfigure}[b]{0.24\textwidth}
       \includegraphics[width=\textwidth]{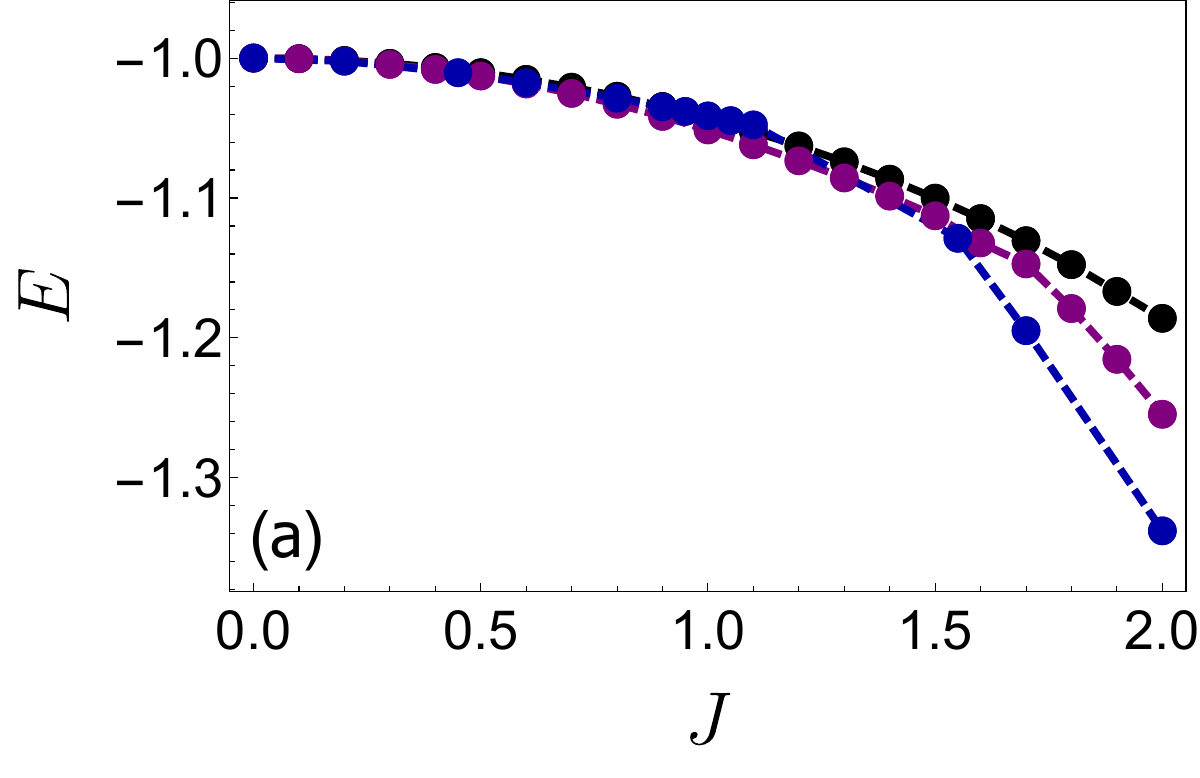}
    \end{subfigure}
    \begin{subfigure}[b]{0.24\textwidth}
       \includegraphics[width=\textwidth]{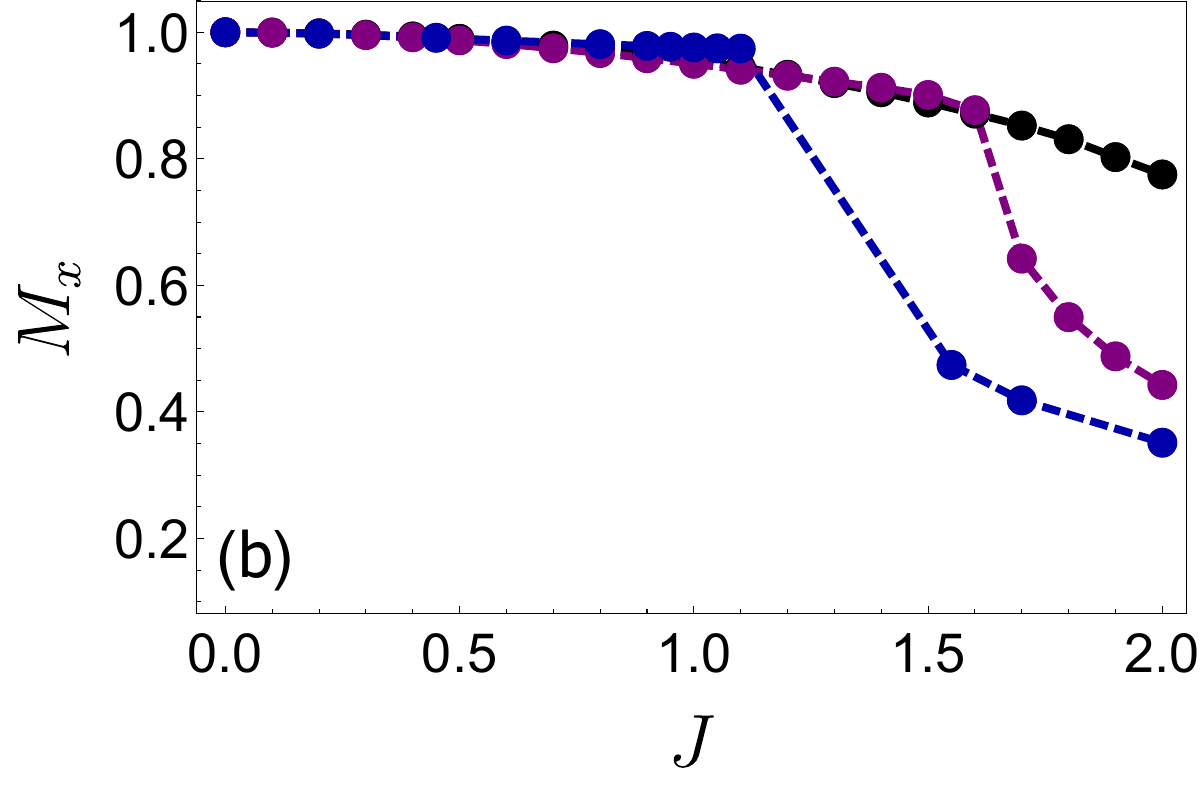}
    \end{subfigure}
    \begin{subfigure}[b]{0.24\textwidth}
     \includegraphics[width=\textwidth]{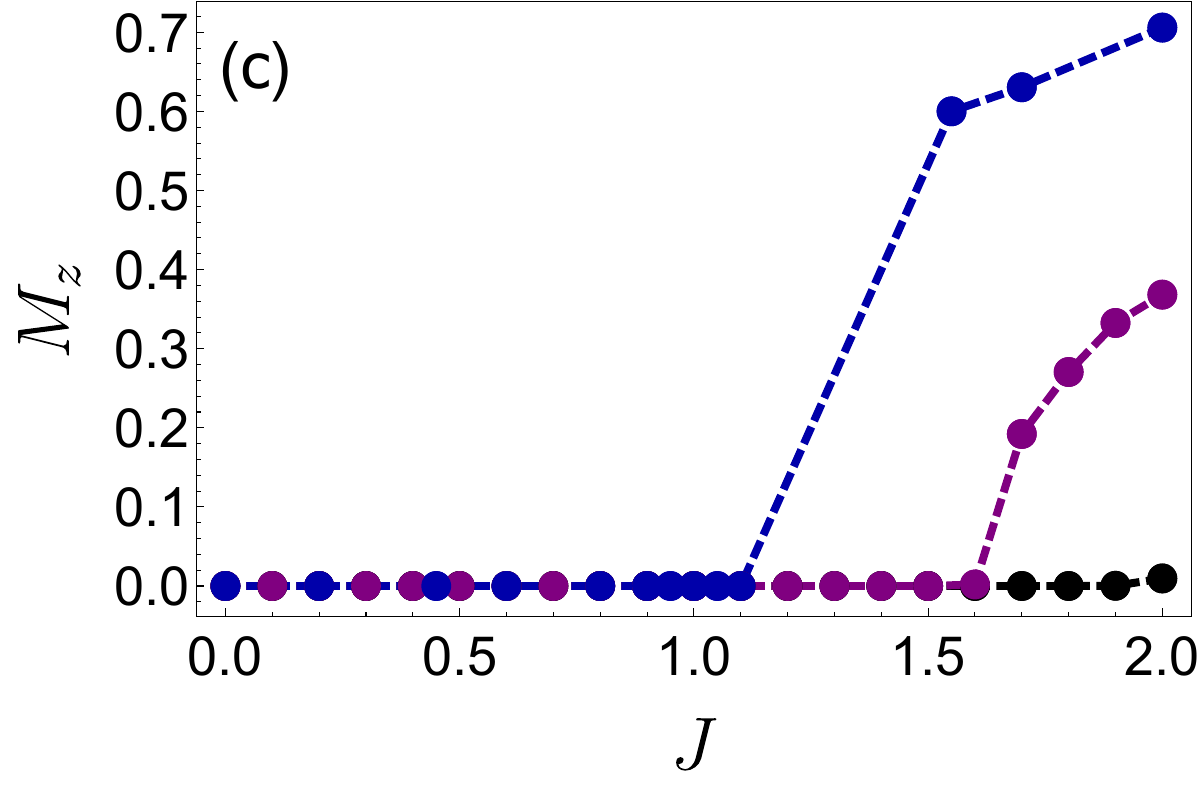}
  \end{subfigure}
  \begin{subfigure}[b]{0.24\textwidth}
     \includegraphics[width=\textwidth]{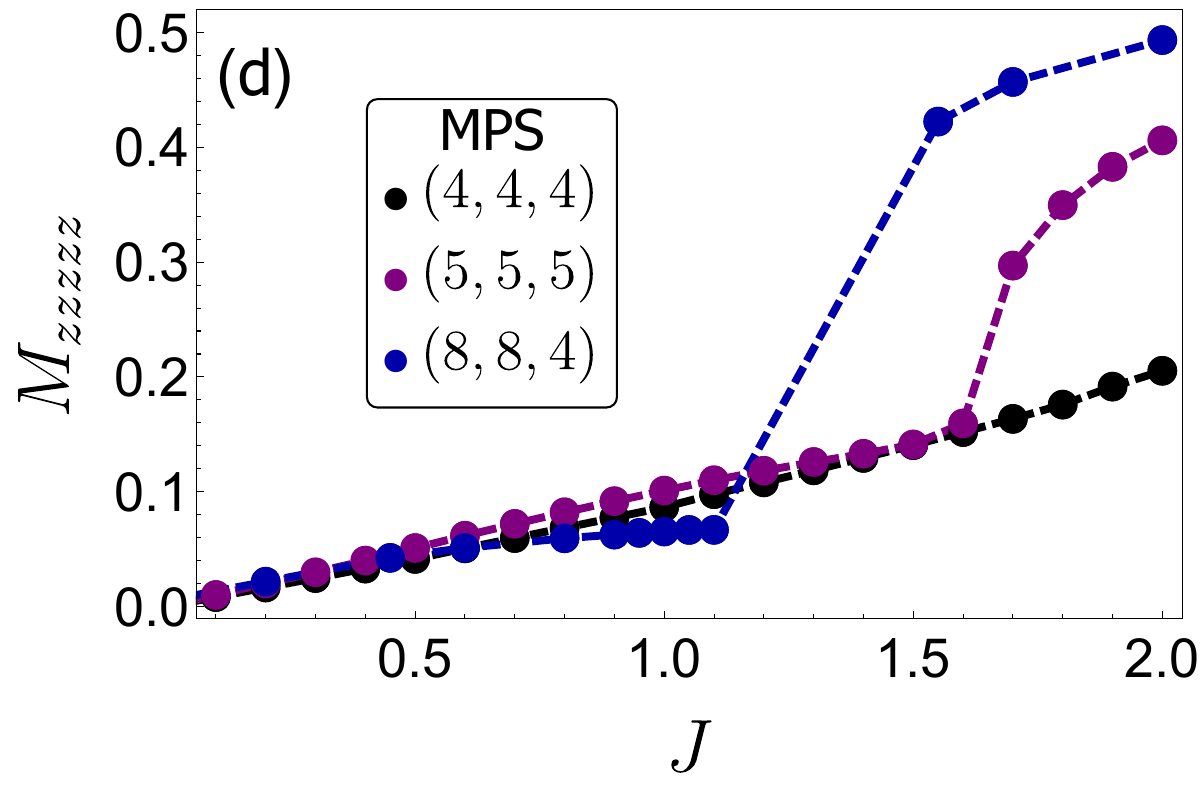}
  \end{subfigure}
    \caption{{\bf Quantum phase transition of $\opcatSPyM{H}$ for OBC.}
    (a) Ground state energy, $E$, per unit size as a function of $J$, for system sizes $L \times L \times M$.
    (b-d) The transverse and longitudinal magnetizations and the average five-spin interaction, respectively. 
    }
    \label{fig:MPS_SPYM_OBC}
\end{figure*}

\begin{figure*}
   \centering
   \begin{subfigure}[b]{0.3\textwidth}
       \includegraphics[width=54mm, height=39mm]{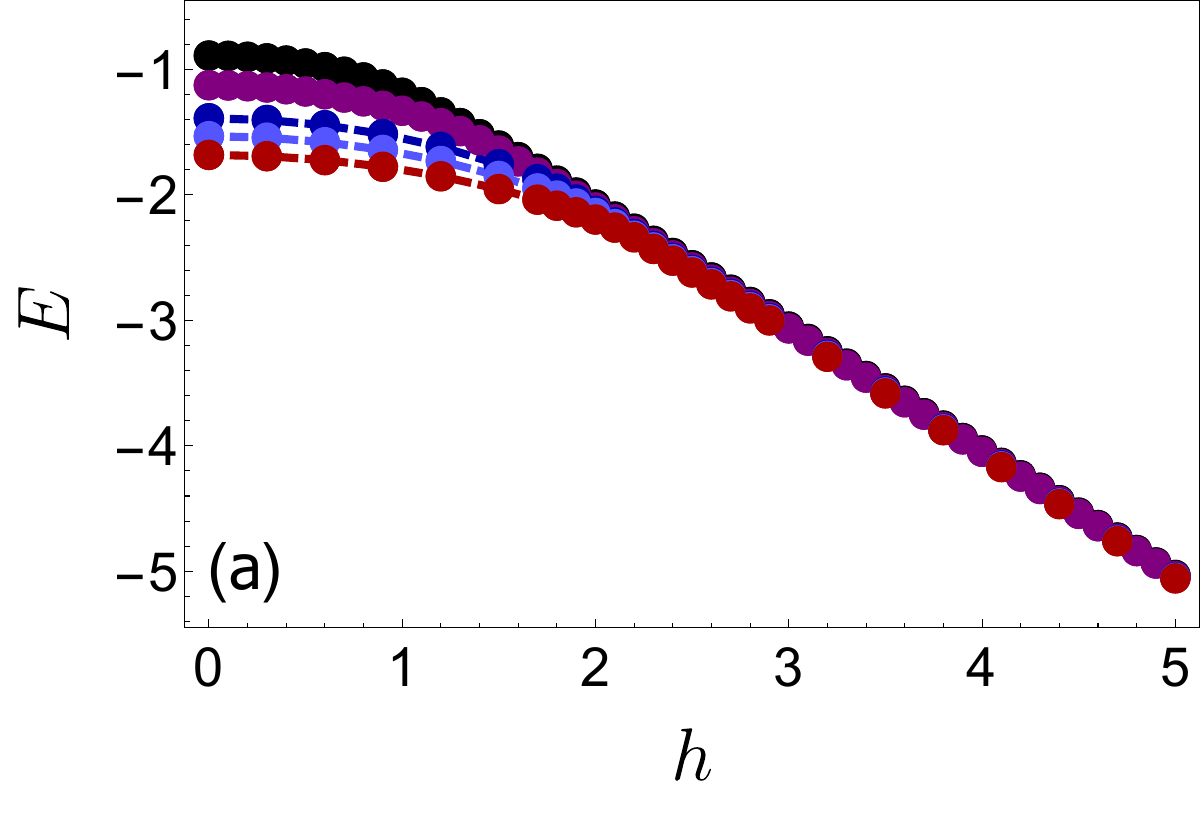}
    \end{subfigure}
    \begin{subfigure}[b]{0.3\textwidth}
       \includegraphics[width=54mm, height=39mm]{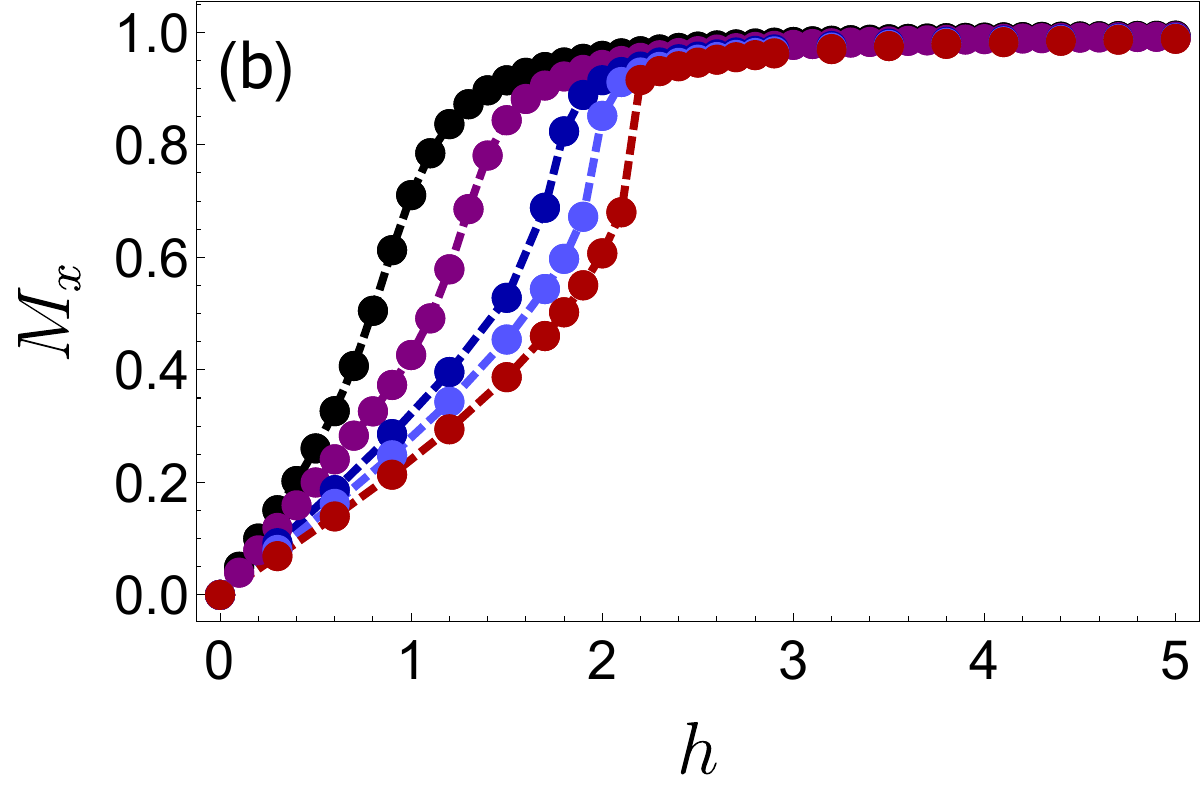}
    \end{subfigure}
    \begin{subfigure}[b]{0.3\textwidth}
      \includegraphics[width=54mm, height=39mm]{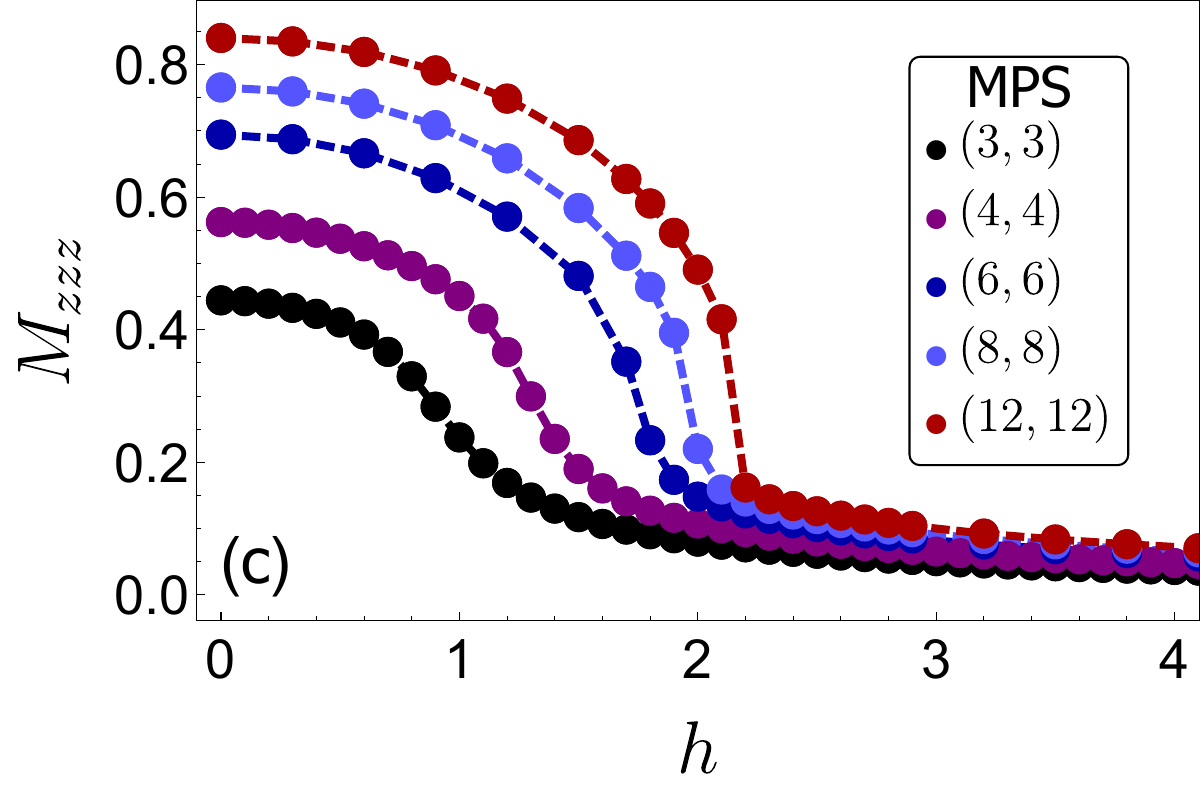}
   \end{subfigure}
   \begin{subfigure}[b]{0.3\textwidth}
     \includegraphics[width=54mm, height=39mm]{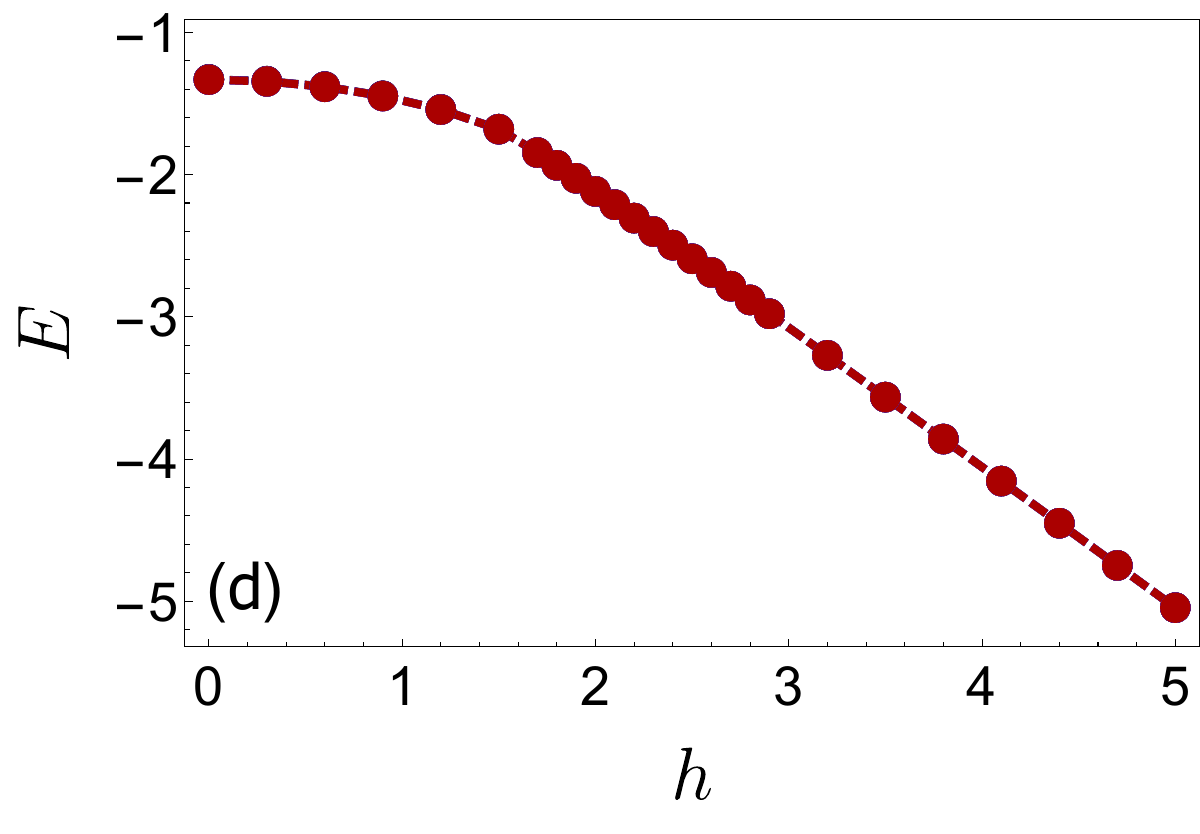}
  \end{subfigure}
  \begin{subfigure}[b]{0.3\textwidth}
     \includegraphics[width=54mm, height=39mm]{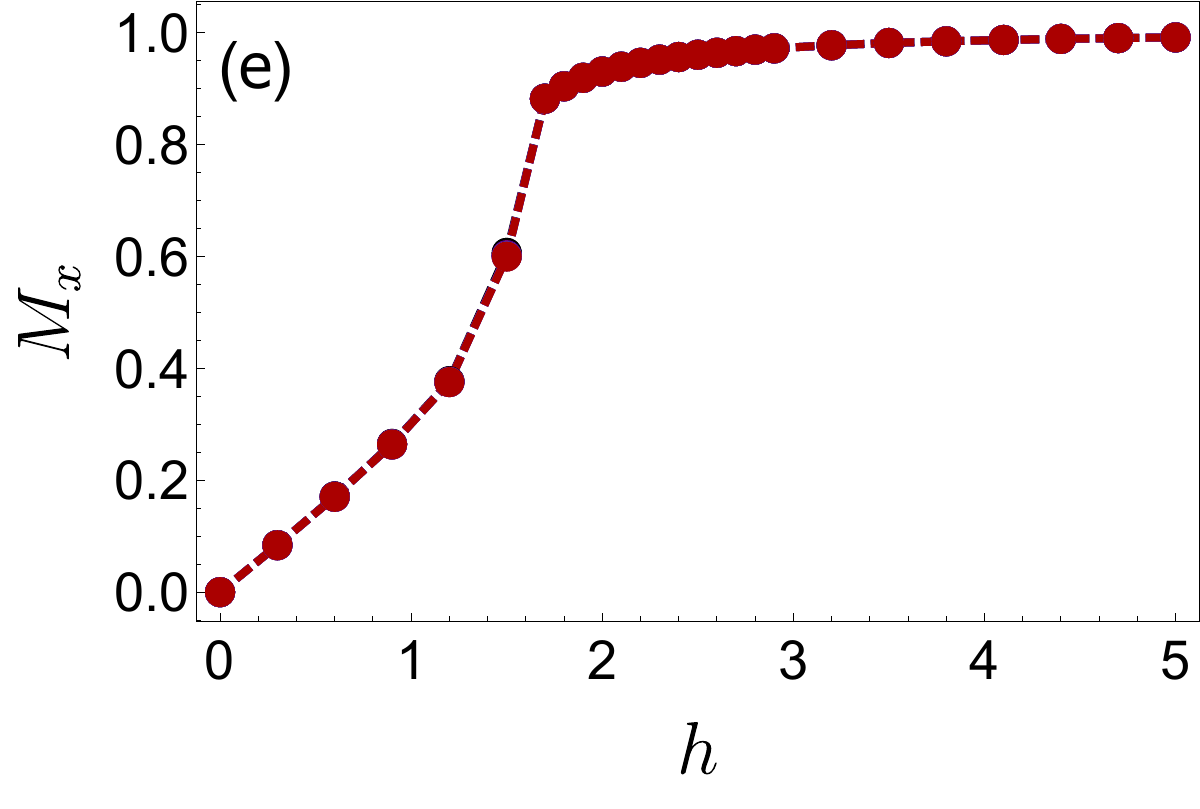}
  \end{subfigure}
  \begin{subfigure}[b]{0.3\textwidth}
    \includegraphics[width=54mm, height=39mm]{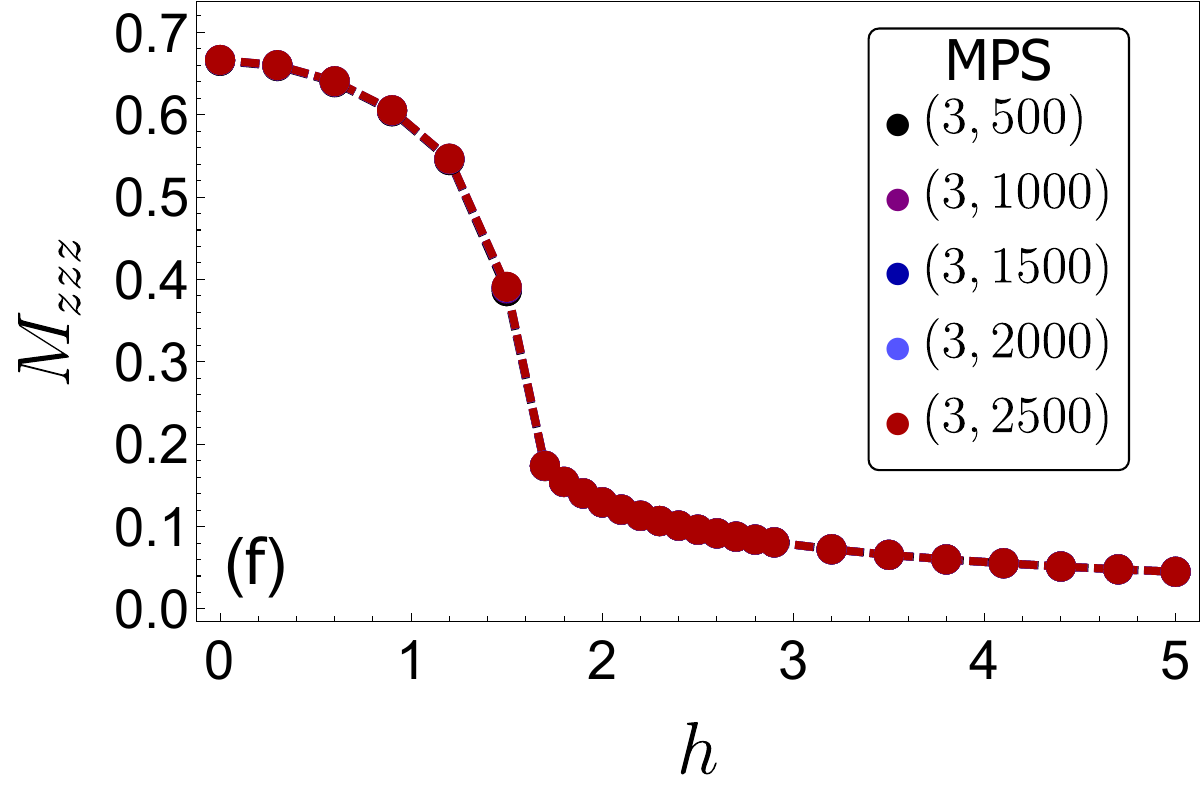}
 \end{subfigure}
   \caption[BW, OBC, square]{
     {\bf Quantum phase transition of the qBW model for OBC.}
     (a) The normalized by the system size ground state energy as a function of $h$. 
     (b,c) Transverse magnetization and three-spin correlator for the interaction of Rule 60 as a function of $h$.
     (d-f) Similar to (a-c) but for the rectangular system sizes given in the inset of panel (f).}
    \label{fig:BW_MPS_OBC}
\end{figure*}

Here we present similar numerical results for the quantum spin models in the case of open boundary conditions. Results are shown in Fig.~\ref{fig:Rule_150_OBC} for $H_{150}$, in Fig.~\ref{fig:MPS_SPYM_OBC} for $\opcatSPyM{H}$, and in Fig.~\ref{fig:BW_MPS_OBC} for the quantum Baxter-Wu model.

The presence of a quantum phase transition for $H_{150}$ for OBC is obscured, as seen from Fig.~\ref{fig:Rule_150_OBC}. The existence of a subextensively large number of classical ground states, which correspond to low-lying states for the quantum model, is expected to reduce the accuracy of the MPS simulations close to the phase transition (if any exists). For strip geometries there is no indication of a discontinuity. This situation resembles the one encountered for OBC for the TPM in Ref.~\cite{sfairopoulos2023boundary}. For the 3D quantum SPyM with OBC, see Fig.~\ref{fig:MPS_SPYM_OBC}, results are even less accurate due to the intrinsic limitations of our MPS simulations in 3D and the very restricted system sizes which are accessible.

For the quantum Baxter-Wu model with OBC, Fig.~\ref{fig:BW_MPS_OBC}, we observe a different behaviour from that of the models above. The major discrepancy with PBC is that the transition has not converged to the value $h=2.4$ yet (although this convergence is not guaranteed either). Apart from this issue, numerics seem to suggest that the phase transition for the quantum Baxter-Wu model with OBC is also of first-order.

\section{Low-lying energy spectra for Rule 150}{\label{appendix:low-lying}}

\begin{figure*}
    \centering
     \begin{subfigure}[b]{0.3\textwidth}
        \includegraphics[width=54mm, height=39mm]{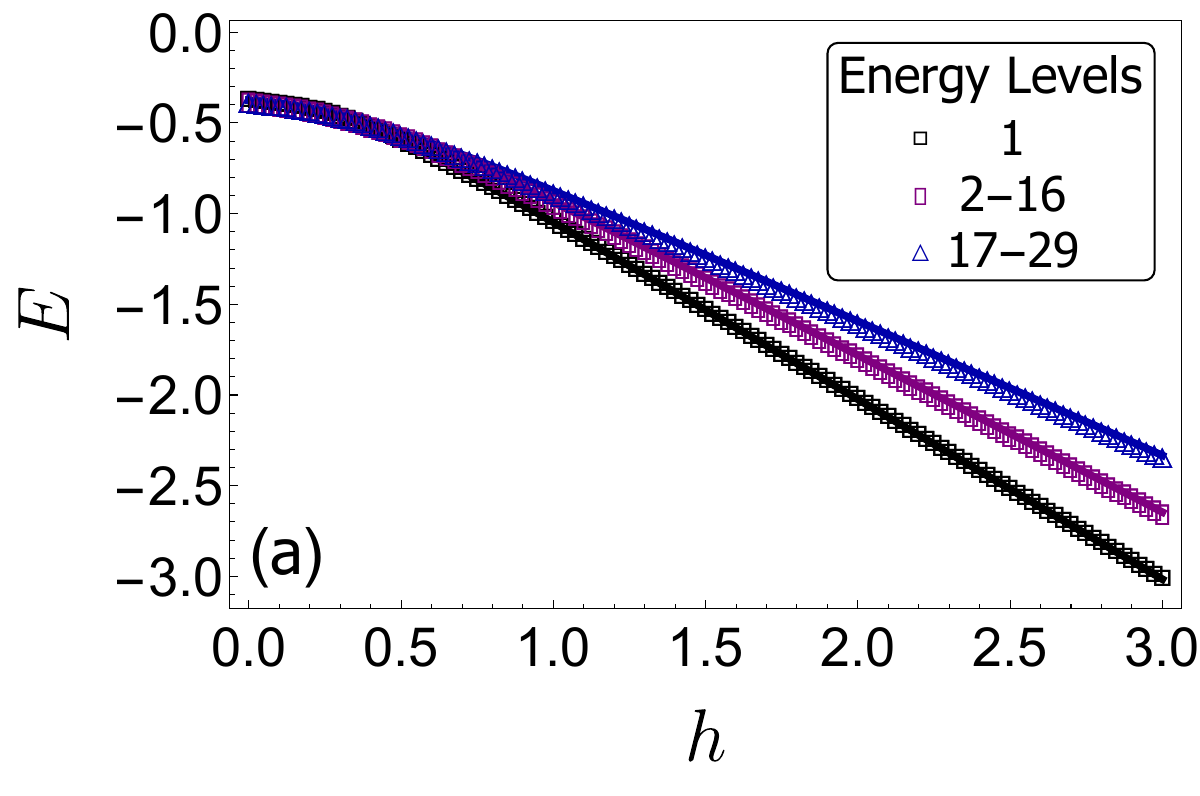}
     \end{subfigure}
     \begin{subfigure}[b]{0.3\textwidth}
        \includegraphics[width=54mm, height=39mm]{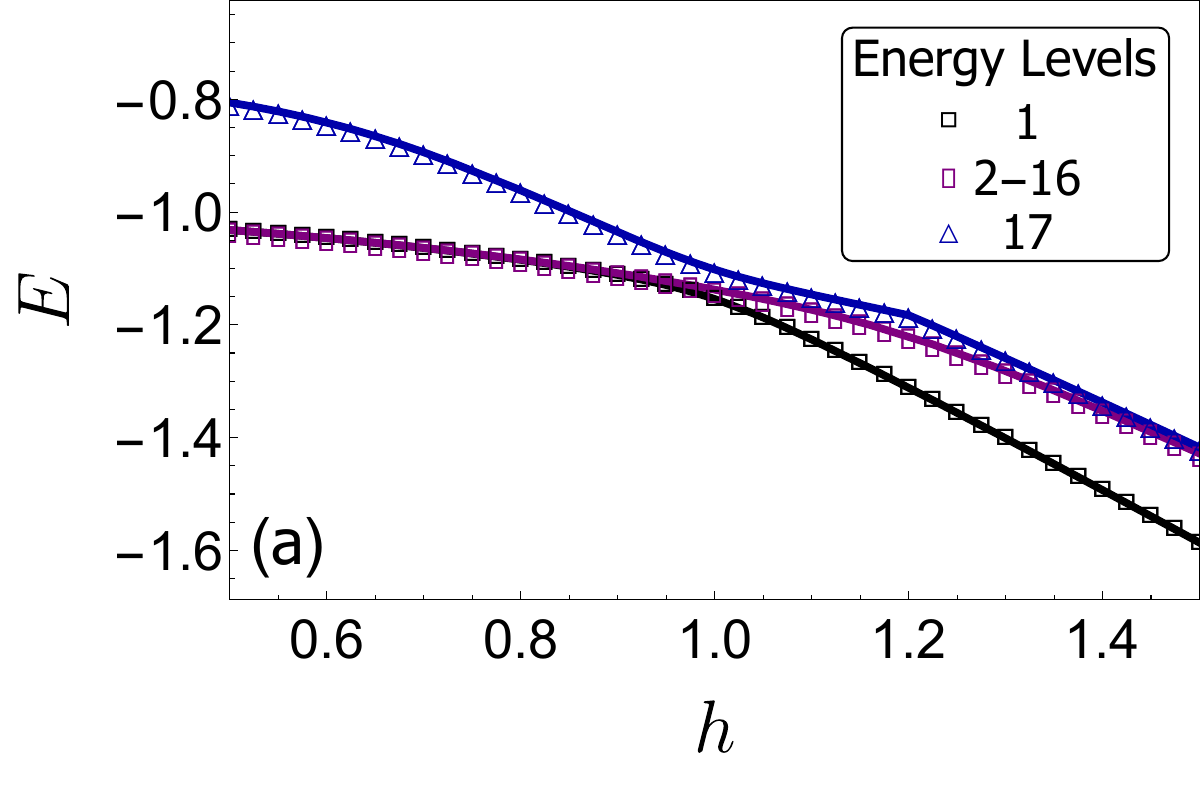}
     \end{subfigure}
     \begin{subfigure}[b]{0.3\textwidth}
        \includegraphics[width=54mm, height=39mm]{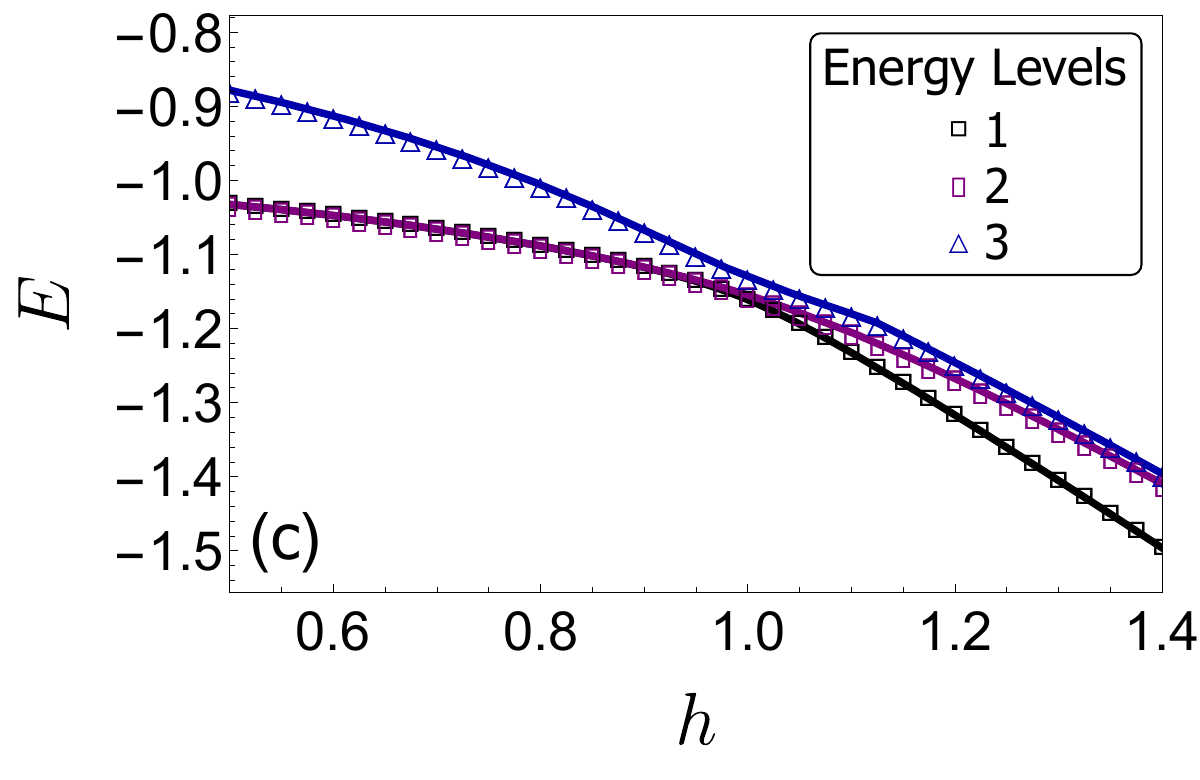}
     \end{subfigure}
     \caption{
        {\bf Low-lying spectrum of $H_{150}$.} 
        (a) Ground state (level 1) and the first 28 excited states (levels 2-29) as a function of $h$ for a system size $4 \times 4$ for OBC. As seen from this panel, the avoided gap crossing is not formed by any of the low-lying states, but involves a higher excited state. 
        (b) Ground state (level 1) and the first 16 excited states (levels 2-17) as a function of $h$ for a system size $4 \times 4$ for PBC. 
        The avoided crossing is indicative of a first-order transition in the ground state, while the splitting of excited levels that of SSB.             
        (c) Same with (b) for the first 3 levels of a system size $3 \times 6$ for PBC.    
     }
     \label{fig:statediagram150}
\end{figure*}

Figure~\ref{fig:statediagram150} shows the low-lying spectrum for two system sizes tractable via ED for $H_{150}$. In panel 
(b) we show the case of size $4 \times 4$ for PBC: here we show the 16 lowest energy configurations given by the two fixed points (the all up and all down states) and the 12 period 2 cycles of the CA, see Table~\ref{tab:periods_150}; the avoided crossing at $J=1.0$ is the finite size signature of the eventual first-order transition, while the splitting away from the ground state of levels 2 to 16 for $h>1$ is a signature of spontaneous breaking of the symmetries. In panel (c) we show the same for a system of size $3 \times 6$ where there are no limit cycles and the only classical minima are the all up and all down states; here it is enough to show the ground state and the first 2 excited states. 
Similarly to the PBC case, the energy spectrum from ED for the lowest energy states is presented for OBC in panel (a).
Similar behaviour to the PBC is observed. However, the state space involved in the SSB involves many more states ($2^{L + 2M - 2}$ specifically for $H_{150}$) and results in the avoided level crossing which would signal the first-order behaviour of the transition to be less pronounced and related to a higher energy excited state.

\bibliography{bibliography-18032025}

\end{document}